\documentclass[aps,prx,twocolumn,superscriptaddress,longbibliography,nofootinbib,floatfix]{revtex4-2}

\usepackage{amsmath,amssymb}
\usepackage{graphicx}
\usepackage{bm}
\usepackage[colorlinks=true,linkcolor=blue,citecolor=blue,urlcolor=blue]{hyperref}

\graphicspath{{./}}

\newenvironment{result}[1]{\par\medskip\noindent\textbf{\textit{Result #1.}}\ \itshape}{\par\medskip}

\newenvironment{theorem}[1]{\par\medskip\noindent\textbf{Theorem #1.}\ \itshape}{\par\medskip}
\newenvironment{lemma}[1]{\par\medskip\noindent\textbf{Lemma #1.}\ \itshape}{\par\medskip}
\newenvironment{proposition}[1]{\par\medskip\noindent\textbf{Proposition #1.}\ \itshape}{\par\medskip}
\newenvironment{corollary}[1]{\par\medskip\noindent\textbf{Corollary #1.}\ \itshape}{\par\medskip}
\newenvironment{remark}{\par\medskip\noindent\textit{Remark.}\ }{\par\medskip}
\newenvironment{proof}[1][Proof.]{\par\noindent\textit{#1}\ }{\hfill$\square$\par\medskip}

\newcommand{\ket}[1]{|#1\rangle}
\newcommand{\bra}[1]{\langle #1|}
\newcommand{\braket}[2]{\langle #1|#2\rangle}

\begin{document}

\title{The exact information cost of randomized measurements in quantum metrology}

\author{Wanqing Li}
\affiliation{School of Instrumentation and Optoelectronic Engineering, Beihang University, Haidian, Beijing 100191, China}

\author{Jiaxin Liu}
\email{matting@buaa.edu.cn}
\affiliation{Hangzhou Institute of Extremely-Weak Magnetic Field Major National Science and Technology Infrastructure, Hangzhou 310051, China}
\affiliation{School of Instrumentation and Optoelectronic Engineering, Beihang University, Haidian, Beijing 100191, China}

\date{October 31, 2026}

\begin{abstract}
Randomized measurements provide state-independent readout for multiparameter quantum sensing. Although 3-designs retain a constant fraction of the quantum Fisher information (QFI) for every pure state, we show that 2-designs need not. In dimension $d$, complete mutually unbiased bases (MUBs) retain exactly $1/(d+1)$ of the QFI in the worst state and direction, attaining the universal lower bound for 2-designs built from orthonormal bases; symmetric informationally complete (SIC) measurements retain exactly $(1-1/\sqrt{d+1})/d$. The loss comes from aligned blind spots. Relative to a basis, pure-state changes split into population and phase parts of equal dimension, but probabilities respond only to the former. Mutual unbiasedness turns a population change for one basis into a phase change for every other. The worst case already occurs for a qubit encoded in two levels. Independent random bases instead scatter their blind spots. In sufficiently large dimensions, five Haar-random bases retain, with high probability, a dimension-independent QFI fraction for every pure state and direction. Random Clifford measurements on any number of qubits retain exactly one third. Thus blind-spot geometry, not second-order symmetry alone, controls worst-case metrological performance.
\end{abstract}

\maketitle

\section{Introduction}
\label{sec:intro}

A quantum sensor encodes unknown parameters in a quantum state and reads them out by measurement~\cite{Giovannetti2011,Degen2017,Pezze2018,Montenegro2025}. The Fisher information of the measurement fixes how many repetitions reach a given precision, and its largest value over all measurements is the quantum Fisher information (QFI)~\cite{Holevo2011,Braunstein1994}. The standard strategy is to choose the measurement that attains the QFI for the state the sensor prepares, as Fisher-symmetric measurements do for pure states~\cite{Li2016} and adaptive schemes do with a growing estimate of the state~\cite{Vargas2024,Martinez2025}. With several parameters the optimal measurements for different parameters conflict~\cite{Ragy2016,Demkowicz2020,Candeloro2024,Chen2024npj}, and recent experiments with entangled atoms, trapped ions, spin defects and photons~\cite{Li2026array,Valahu2025,Isogawa2026,Mi2025} balance the conflicting measurements with one measurement designed for their state. The strategy requires the state to be known before the measurement is chosen. The state, however, carries the unknown parameters, so the measurement is designed for a prior estimate and has to be redesigned whenever the state or the parameters of interest change.

A randomized measurement takes the opposite strategy. A random unitary, drawn without regard to the state or to the quantities of interest, rotates the state before a fixed projective measurement, and the experimenter records both the unitary and the outcome~\cite{Elben2023toolbox,Cieslinski2024}. One data set, taken on single copies, therefore serves every parameter, and the same records predict many observables from few repetitions~\cite{Huang2020shadows,Grier2024pure,Helsen2023thrifty,Vermersch2024}, reveal entanglement and many-body dynamics~\cite{Brydges2019,Wyderka2023,Joshi2024}, estimate nonlinear properties and the QFI~\cite{DuTang2026,Rath2021QFI,Vitale2024QFI}, learn Hamiltonians~\cite{Huang2023Hamiltonian,Dutkiewicz2024} and certify quantum states~\cite{Huang2024certify,Gupta2025certify}. Randomized-measurement protocols run on trapped ions, photons and superconducting circuits~\cite{Brydges2019,Joshi2024,Lib2025photonic,Hu2025shallow}, with proposals for Rydberg arrays~\cite{Notarnicola2023}. Shallow random circuits~\cite{Bertoni2024shallow,Schuster2025lowdepth,Du2026} or the native dynamics of analog simulators~\cite{Choi2023,Cotler2023,McGinley2023} supply the randomness cheaply. In exchange, a randomized measurement retains only part of the QFI. How large a part depends on the direction in which the parameters move the state, because the Fisher information of a measurement can spread unevenly over the directions of a state~\cite{Saini2026,AitHaddou2026,Taherpour2026}. For a pure state the spread has a hard limit, because a change and its conjugate change share one information budget, which every measurement must divide between them~\cite{Gill2000,Matsumoto2002,Lu2021,Yung2025,Wang2026}. No single-copy measurement therefore secures more than half of the QFI for all directions at once, and collective measurements on several copies do no better~\cite{Matsumoto2002,Demkowicz2020}. A randomized measurement acts on single copies, so its quality is the fraction of the QFI it retains in the worst direction, and by this measure the best randomization is a Haar-random basis, drawn uniformly at random, which retains exactly half in every direction of every pure state~\cite{Hayashi1998,Li2016,Lu2025}. Exact Haar randomness is costly, but finite ensembles keep a guarantee when they reproduce Haar averages up to third order~\cite{Mele2024}, since randomization over such a 3-design retains at least a quarter of the QFI for every pure state and any number of parameters~\cite{Zhou2026}.

In practice, the most used randomizations are 2-designs, which reproduce Haar averages up to second order, namely complete sets of mutually unbiased bases (MUBs), symmetric informationally complete measurements (SICs) and random Clifford circuits on qudits of odd prime dimension~\cite{Klappenecker2005,Renes2004,Webb2016,Zhu2024magic}. In a complete set of MUBs, certainty in one basis implies uniformly random outcomes in every other~\cite{Wootters1989,Durt2010}. MUBs are the standard measurements for tomography, key distribution and entanglement certification with high-dimensional photons~\cite{Adamson2010,Cerf2002,Zahidy2024,Bavaresco2018,Morelli2023,Li2025schmidt,Erhard2020,Goel2024}. SICs are the minimal 2-designs and have been realized on trapped-ion processors~\cite{Stricker2022,Brinster2026}. Random Clifford circuits are the native randomizer of quantum processors and form 3-designs on qubits~\cite{Zhu2017}, so the guarantee above covers them. On qudits, which processors and sensors increasingly use~\cite{Wang2020qudits,Ringbauer2022,Meth2025,Low2026,Nguyen2024,Champion2025,Brock2025,Kristen2020,Deng2024,Ahmed2025,Lib2026}, all these randomizations stop at second order~\cite{Anand2026}, and whether 2-designs retain a constant fraction of the QFI has remained open~\cite{Zhou2026}.

Related problems suggest conflicting answers to this question. In learning, stopping at second order costs a factor that grows with the dimension. For classical shadows, every 2-design is optimal for typical observables, while complete MUBs, SICs and qudit Clifford circuits need more repetitions for the worst ones, by a factor proportional to the dimension~\cite{Mao2025,Yang2026}. A single magic gate, the non-Clifford resource behind universal quantum computation and higher designs~\cite{Bravyi2005,Haferkamp2023}, removes this overhead of Clifford circuits~\cite{Mao2025,Zhu2024magic}, whereas worst-case-optimal learning needs a number of bases that grows as the square of the dimension~\cite{Yang2026,Castellano2026}. Certification shows a similar gap between typical and worst-case states~\cite{Huang2024certify,Coladangelo2026,Gupta2025certify}. Because Fisher information and shadow estimation are closely linked~\cite{Cho2026,Chen2026COLT,Mao2026}, sensing might suffer the same loss, by a factor that grows with the dimension. The symmetry of MUBs points the other way, since it embodies Bohr's complementarity, underlies entropic uncertainty relations and the trade-off between which-path knowledge and fringe visibility~\cite{Coles2017,Greenberger1988,Englert1996,Durr2001}, and makes MUBs optimal for tomography~\cite{Wootters1989}. Phase retrieval points to a third answer, a few unstructured bases. A handful of bases determines every pure state in any dimension~\cite{Mondragon2013,Goyeneche2015}, and a constant number of random bases does so stably~\cite{Voroninski2013,Baldwin2016}, as random measurements do in coherent imaging~\cite{Candes2013,Shechtman2015,Dong2023}. Faced with an unknown state of many levels, should one measure in a complete set of perfectly unbiased bases or in a dozen random ones? Design averages cannot decide, because the Fisher information divides the squared response of each outcome by its probability, and the average of a ratio can differ greatly from the ratio of averages.

In this work we show that being a 2-design does not guarantee a constant fraction of the QFI. The reason is that measurement probabilities are insensitive to the phases of a quantum state. Every measurement basis therefore acts like a polarizer that registers changes of populations and is blind to changes of phases, so it resolves exactly half of the ways a pure state can change, and a randomized measurement retains in each direction the average transmission of its bases. Complementarity aligns the blind spots of different bases, because a population change for one basis is a phase change for every basis unbiased to it, so in a complete set of MUBs one basis sees the change and all the others are blind to it. Complete MUBs and random Clifford measurements of a qudit of odd prime dimension therefore lose exactly a factor of the dimension plus one, the worst value among 2-designs built from orthonormal bases, and SICs lose a similar factor. The loss appears on simple states, such as a qubit stored in two levels of a qudit, and averages over random states miss it, because the same measurements are asymptotically optimal on typical states. Unstructured bases are blind to different changes, so scattering the blind spots removes the loss. Five random bases retain a dimension-independent fraction, a dozen outperform complete MUBs from about eleven levels on, a single magic gate restores a constant fraction, and random Clifford measurements of any number of qubits retain exactly one third. Thus, once perfect symmetry is given up, one fixed measurement reaches, for every parameter of every pure state in every dimension, a precision within a constant factor of the quantum limit. For sensors, the results give exact error budgets and a design rule on qubits and qudits, and structured states become the stringent benchmark. For quantum learning, every shadow-norm bound becomes a sensing guarantee, and sensing needs far fewer bases than learning.

\begin{figure*}[t]
\centering
\includegraphics[width=\textwidth]{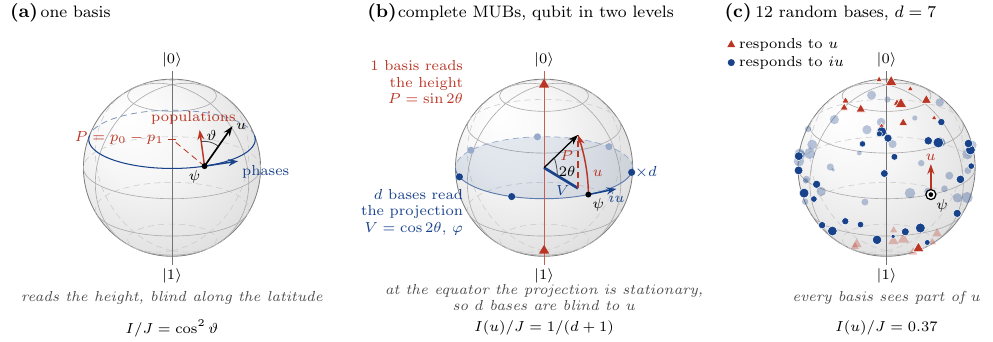}
\caption{\textbf{Mutually unbiased bases are collectively blind.}
(a)~One basis reads the height. Measured in the basis of the poles $\ket0,\ket1$, a qubit state $\psi$ gives probabilities that depend only on the height $P=p_0-p_1$ of its Bloch vector. Changes of populations, along the meridian (red), are seen. Changes of phases, along the latitude circle (blue), are invisible. A change $u$ at angle $\vartheta$ to the population direction is seen with Malus's law, $I/J=\cos^2\vartheta$.
(b)~Complete MUBs on a qubit stored in two levels $\ket0,\ket1$ of a qudit, for $d=7$. The which-level basis has its outcomes at the poles (red triangles) and reads the height $P$. The $d$ interference bases have their outcomes at $d$ equally spaced points of the equator (blue circles, $\times d$) and read the projection of the Bloch vector onto the equatorial plane. The projection has the fringe visibility $V$ as its length and the phase $\varphi$ as its direction. The tilt $u$ raises the Bloch vector of $\psi=(\ket0+\ket1)/\sqrt2$ by an angle $2\theta$, so the height $P=\sin2\theta$ changes at first order and the projection $V=\cos2\theta$ only at second order. All interference bases are blind to $u$, and $I(u)/J=1/(d+1)$.
(c)~Twelve random bases at $d=7$. Each outcome is projected onto the encoded qubit and drawn at its Bloch direction, with marker size growing with its weight. Faint markers lie on the far hemisphere. Outcomes near the poles respond mainly to $u$ (red triangles), and outcomes near the equatorial plane mainly to $iu$ (blue circles). Every basis sees part of $u$, and $I(u)/J=0.37$. The same pictures hold for any state and direction, with the qubit spanned by $\psi$ and $u$.}
\label{fig:mechanism}
\end{figure*}

\section{Fisher information of randomized measurements}
\label{sec:half}

Consider a pure state $\ket{\psi_\theta}$ of a $d$-level system, or qudit, that depends on unknown parameters $\theta$, and a measurement repeated on many copies. A small change of the parameters moves the state along a direction $u$. We take $u$ orthogonal to $\psi$, because the component along $\psi$ only changes the global phase. A measurement with outcome probabilities $p_s$ then has the Fisher information and the QFI
\begin{equation}
I(u)=\sum_s\frac{(\partial_u p_s)^2}{p_s},\qquad J(u)=4\|u\|^2 ,
\label{eq:fisher}
\end{equation}
where $\partial_u p_s$ is the response of outcome $s$ to the change. The Fisher information measures how sharply the outcome statistics respond, and $N$ repetitions estimate a parameter with variance at best $1/[NI(u)]$. The QFI $J(u)$ is the largest Fisher information that any measurement can extract along $u$. With several parameters, each parameter moves the state along its own direction, $I$ and $J$ become matrices, and one measurement has to share its sensitivity among the directions.

A randomized measurement chooses basis $a$ with probability $\lambda_a$, measures in it, and records both $a$ and the outcome $x$. The outcome $s=(a,x)$ then has probability $p_{a,x}=\lambda_a|\braket{b_{a,x}}{\psi}|^2$, where $\ket{b_{a,x}}$ is the $x$th vector of basis $a$. We measure the quality of a randomized measurement by the retained fraction
\begin{equation}
c=\inf_{\psi,u}\frac{I(u)}{J(u)} ,
\label{eq:retained}
\end{equation}
the worst ratio over all pure states and all directions. A measurement with retained fraction $c$ satisfies $I\succeq cJ$ for every pure-state model. It needs at most $1/c$ times more repetitions than the quantum limit set by the QFI, for every parameter at once. We call it near-optimal when $c$ stays constant as the dimension $d$ grows.

Seen from a single orthonormal basis $\{\ket x\}$, every change of a pure state has two parts. Write the amplitudes of the state as $\braket{x}{\psi}=r_xe^{i\phi_x}$ and those of the change as $\braket{x}{u}=e^{i\phi_x}(\alpha_x+i\beta_x)$, with real $\alpha_x$ and $\beta_x$. The part $\alpha$ changes the moduli $r_x$, that is, the populations of the basis states, and the part $\beta$ changes their phases. The probabilities $p_x=r_x^2$ respond to the first part and ignore the second, $\partial_up_x=2r_x\alpha_x$. For a state with nonzero amplitudes this gives $I_B(u)=4\|\alpha\|^2$ and $J(u)=4(\|\alpha\|^2+\|\beta\|^2)$, so that
\begin{equation}
\frac{I_B(u)}{J(u)}=\cos^2\vartheta_B ,
\label{eq:malus}
\end{equation}
where $\vartheta_B$ is the angle between $u$ and the population changes of the basis. Equation~\eqref{eq:malus} is Malus's law for the changes of a pure state. The basis acts like a polarizer that transmits the population part of a signal and blocks the phase part. We call the phase changes of a basis its blind spot. Population changes and phase changes each fill exactly half of the $2(d-1)$ real directions of change, and multiplying a change by $i$ turns one kind into the other. A single basis therefore sees half of the ways a pure state can change [Fig.~\ref{fig:mechanism}(a)]. For a known state one could rotate the basis until the signal became a pure population change. For an unknown state the right rotation is unknown as well, and the measurement has to be randomized.

Because the basis label is recorded, a randomized measurement averages what its bases see, and its Fisher information is the weighted sum over the bases,
\begin{equation}
\frac{I(u)}{J(u)}=\sum_a\lambda_a\cos^2\vartheta_a .
\label{eq:average}
\end{equation}
The ratio $I(u)/J(u)$ is thus the average transmission of the signal $u$ through a set of polarizers, one of which is chosen at random in each run. If the blind spots point in all directions, as for a Haar-random basis, every signal keeps half of its QFI. This value is the ceiling for all directions at once, because a change $u$ and its conjugate $iu$ share one budget. Each basis gives the fraction $\cos^2\vartheta_a$ to $u$ and $\sin^2\vartheta_a$ to $iu$.

\begin{result}{1 (information budget)}
For every randomized measurement and every pure state whose outcome probabilities are all nonzero, each direction $u$ satisfies
\begin{equation}
I(u)+I(iu)=J(u),\qquad \frac{I(u)}{J(u)}=\frac{1+\eta(u)}{2},
\label{eq:budget}
\end{equation}
with the imbalance $\eta(u)=\sum_a\lambda_a\cos2\vartheta_a$. The same identity holds for every measurement whose outcomes are pure states, including SICs.
\end{result}

\noindent The budget is conserved, so a randomized measurement can only shift information between a change and its conjugate. The worst direction is the one that loses most to its conjugate, and Appendix~\ref{app:budget} gives it at any state. Every loss in this paper, for MUBs, SICs and random Clifford measurements alike, is an imbalance of this kind [Figs.~\ref{fig:mechanism}(b) and \ref{fig:mechanism}(c)]. Result~1 excludes states that make outcomes impossible, for example a state that coincides with a measurement vector. We call such outcomes dark and evaluate the Fisher information at the state itself, where dark outcomes contribute nothing, so the budget can fall short of $J(u)$. Evaluating the Fisher information along a path through the state leaves every worst-case value in this paper unchanged (Appendices~\ref{app:notation} and \ref{app:budget}). Equation~\eqref{eq:budget} is the exact pure-state form of the Gill--Massar bound~\cite{Gill2000,Lu2021}, and its symmetric point $\eta=0$ is the Fisher-symmetric measurement of Ref.~\cite{Li2016}, which is built around a known state.

\begin{result}{2 (known state)}
At a known pure state, every allocation that uses the full budget, $I(u)+I(iu)=J(u)$ for all $u$, is realized by choosing at random between two orthonormal bases. A single basis realizes exactly the extreme allocations, which give the full QFI to half of the directions and nothing to their conjugates.
\end{result}

\noindent Consequently a coin flip between two orthonormal bases of the system itself, with no ancilla, attains the Holevo bound of every pure-state model (Appendix~\ref{app:budget}). This bound is the best precision for any weighting of the parameters~\cite{Holevo2011,Matsumoto2002}. For the symmetric allocation, $I(u)=I(iu)=J(u)/2$, the coin flip is the two-basis construction of Fisher-symmetric measurements~\cite{Li2016}. A known state therefore needs only two bases, and the cost of a randomized measurement comes from not knowing the state.

Without knowledge of the state, a measurement must keep the imbalance small at every state at once. Whether it can depends on how the blind spots of its bases are arranged, and the next section shows that perfect symmetry lines them up.

\section{Mutually unbiased bases and qudit Clifford measurements}
\label{sec:mub}

The simplest place to look for aligned blind spots is a qubit encoded in two levels of a qudit. This configuration is common in experiments, for example two clock or hyperfine levels of a multilevel ion or atom, two modes of a high-dimensional photon, or qubits stored in the levels of a single ion or nuclear spin~\cite{Ringbauer2022,Erhard2020,Campbell2022polyqubit,Low2026,Yu2025}. Take the equal-weight superposition $\psi=(\ket0+\ket1)/\sqrt2$ of two levels. Its population change $u=(\ket0-\ket1)/\sqrt2$ tilts the weights of the two levels, and the conjugate change $iu$ shifts their relative phase. The standard complete set of $d+1$ MUBs contains the basis of the qudit levels~\cite{Wootters1989,Durt2010}, so its bases come in two kinds [Fig.~\ref{fig:mechanism}(b)]. The which-level basis contains $\ket0$ and $\ket1$. It asks which level is occupied and reads the population contrast $P=p_0-p_1$. Each of the other $d$ bases is unbiased to both levels. It recombines them and records interference across its $d$ outcomes, from which one reads the fringe visibility $V$ and the relative phase $\varphi$. We call these $d$ bases the interference bases.

Every basis unbiased to this pair is an interference measurement. Tilt the populations along $\psi_\theta=\cos(\pi/4-\theta)\ket0+\sin(\pi/4-\theta)\ket1$. An outcome $x$ of an unbiased basis $a$ overlaps $\ket0$ and $\ket1$ with equal moduli $1/\sqrt d$ and some relative phase $\chi_{a,x}$, so its probability is
\begin{equation}
p_{a,x}(\theta)=\frac{\lambda_a}{d}\bigl[1+\cos(2\theta)\cos\chi_{a,x}\bigr].
\label{eq:fringe}
\end{equation}
The $d$ outcomes sample an interference fringe at $d$ phases [Fig.~\ref{fig:mechanism}(b)]. The fringe has visibility $V=\cos2\theta$, while the which-level basis reads $P=\sin2\theta$. A pure state obeys $P^2+V^2=1$~\cite{Greenberger1988,Englert1996}, so at equal populations, where $P=0$ and $V=1$, the visibility is stationary. A tilt by $\theta$ therefore changes $P$ at first order and $V$ only at second order, while a phase change of the same size shifts the fringe and changes the probabilities at first order. This asymmetry is wave--particle duality. Every outcome of every unbiased basis inherits the stationarity of $V$, so to all $d$ interference bases the population change is a pure phase change, and the which-level basis alone sees it. Away from equal populations $V$ responds linearly to the tilt, and the interference bases regain sensitivity (Sec.~\ref{sec:design}).

\begin{result}{3 (collective blindness)}
Mix any $K$ mutually unbiased bases with weights $\lambda_a$. For the encoded qubit built from two vectors of basis $a$, the population change retains $I(u)/J(u)=\lambda_a$. Hence
\begin{equation}
c_{K\,\mathrm{MUB}}\le\min_a\lambda_a\le\frac1K .
\label{eq:mub}
\end{equation}
For the standard MUBs of odd prime dimension $d$, the conjugate phase change of $(\ket0+\ket1)/\sqrt2$ retains $I(iu)/J(u)=d/(d+1)$.
\end{result}

\noindent For complete MUBs with equal weights the ceiling is $1/(d+1)$, and Result~4 below shows that no state does worse, so $c_{\mathrm{MUB}}=1/(d+1)$. By the budget~\eqref{eq:budget}, the which-level basis gives its whole share to $u$, and every interference basis gives its whole share to $iu$. More unbiased bases make the imbalance worse, because every added basis joins the interference side and the ceiling $1/K$ falls.

Because the Cram\'er--Rao bound covers every unbiased estimator, the blindness is a property of the measurement data, and the factor $d+1$ applies to any way of processing the data. The loss also survives when every outcome can occur. After a perturbation of the encoded qubit that makes all $d(d+1)$ outcome probabilities positive, the perturbed state still has $I(u)/J(u)$ within a relative amount of about $4/d^2$ of $1/(d+1)$ (Appendix~\ref{app:mub}).

The blindness extends to random Clifford measurements, the most common qudit randomizer. A random Clifford measurement of one qudit applies a uniformly random Clifford unitary and then measures in the computational basis. For odd prime $d$, the $d(d+1)$ stabilizer states of a qudit form a complete set of MUBs, and a uniformly random Clifford unitary carries any of them to each of them with equal frequency. The measurement therefore has the outcome statistics of complete MUBs and loses a factor $d+1$ (Appendix~\ref{app:mub}). On $n$ qudits of prime dimension $p$, an encoded qubit on one qudit, with the others in $\ket0$, retains $1/(p+1)$ for every $n$.

Complementarity therefore has a metrological cost. Unbiased bases ignore the same change, since a population change for one basis is a phase change for all the others. At equal populations, a population-type parameter of an encoded qubit, such as a Rabi angle, a beam-splitter ratio or a transition probability, needs $d+1$ times more repetitions than the quantum limit, twelvefold at $d=11$.

\section{Bounds for 2-designs}
\label{sec:worst}

\begin{result}{4 (extremality)}
(a)~Every 2-design built from orthonormal bases retains at least $1/(d+1)$ of the QFI, and complete MUBs attain this value. (b)~Every 2-design retains at least half of this value,
\begin{equation}
c\ge\frac1{d+1}\ \ \text{(built from bases)},\qquad c\ge\frac1{2(d+1)}\ \ \text{(all)}.
\label{eq:extremal}
\end{equation}
\end{result}

\noindent The bound (a) comes from a zero-sum rule. Within one basis, probability that leaves one outcome must arrive at the others, so a basis that responds to a change moves at least two outcomes, in opposite directions. This rule guarantees that a responding basis collects Fisher information at least twice the sum of its squared responses (Appendix~\ref{app:extremal}). The 2-design property fixes this sum, averaged over the bases, to its Haar value. Together the two facts give $1/(d+1)$. Equality requires every basis but one to be blind and the remaining basis to move exactly two outcomes, which is the MUB pattern of Fig.~\ref{fig:mechanism}. For a general 2-design the squared responses alone give bound (b), which is half as large as bound (a).

A SIC, the other maximally symmetric measurement, is a 2-design without basis structure, made of $d^2$ pure states with equal pairwise overlaps~\cite{Renes2004}. Every other SIC vector overlaps any two SIC vectors with the same modulus, so it is unbiased to the pair, and the blindness of Sec.~\ref{sec:mub} reappears.

\begin{result}{4(c) (SICs)}
Every SIC measurement retains exactly
\begin{equation}
c_{\mathrm{SIC}}=\frac{1-1/\sqrt{d+1}}{d},
\label{eq:sic}
\end{equation}
attained by the in-phase superposition of two SIC vectors along its population change.
\end{result}

\noindent The value lies below $1/(d+1)$ because the two SIC vectors overlap with modulus $1/\sqrt{d+1}$, and the two values agree for large $d$. The two most symmetric measurements of quantum theory thus sit within a factor of two of the universal floor $1/[2(d+1)]$. Second moments guarantee a retained fraction of order $1/d$, and this order is sharp.

With third moments the loss disappears, since every 3-design retains at least $(d+2)/[4(d+1)]\ge1/4$~\cite{Zhou2026}. The qubit is special, because for $d=2$ the complete MUBs are the three Pauli bases, which form a 3-design.

\begin{result}{4(d) (qubit)}
For a qubit, Pauli measurements satisfy $1/3\le I(u)/J(u)\le2/3$ for every state and direction, so the 3-design bound is tight at $d=2$.
\end{result}

\begin{figure}[t]
\centering
\includegraphics[width=\columnwidth]{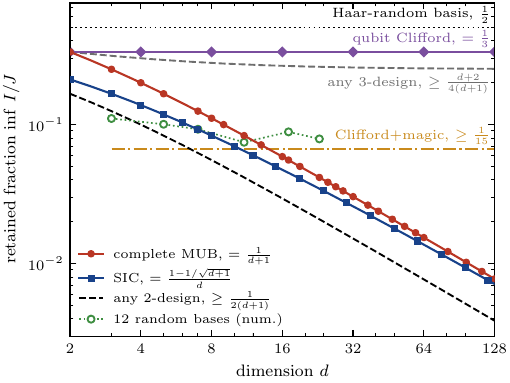}
\caption{\textbf{Retained fraction of randomized measurements.} Retained fraction $\inf_{\psi,u}I/J$ versus Hilbert-space dimension $d$. Solid lines and filled markers show exact values, $1/(d+1)$ for complete MUBs (markers at prime powers), $(1-1/\sqrt{d+1})/d$ for SICs and $1/3$ for random Clifford measurements of qubits (markers at $d=2^n$). Dashed lines show proven lower bounds, $1/[2(d+1)]$ for any 2-design, $(d+2)/[4(d+1)]$ for any 3-design~\cite{Zhou2026} and $1/15$ for Clifford circuits with a single magic gate (odd prime-power $d$). The dotted line marks the Fisher-symmetric value $1/2$ of a Haar-random basis. Open circles show twelve independent random bases (adversarial search, minimum over random draws), which overtake complete MUBs near $d\approx11$.}
\label{fig:price}
\end{figure}

\noindent The qubit is the one dimension in which unbiasedness costs nothing extra. From $d=3$ on, the cost of unbiased measurements grows linearly with $d$. In Fig.~\ref{fig:price} the retained fractions of the aligned designs slide down as $1/d$, while those of the measurements that scatter blind spots stay flat.

\section{Typical and structured states}
\label{sec:averages}

Averages over random states miss this loss, because the 2-design property enforces a cancellation that survives only when all outcomes have comparable overlaps with the state. Write a measurement whose outcomes are pure states with unit vectors $\ket s$ and weights $w_s$. For a randomized measurement, $s=(a,x)$, $\ket s=\ket{b_{a,x}}$ and $w_s=\lambda_a$. For each outcome define the response amplitude $z_s=\braket{\psi}{s}\braket{s}{u}$ and the overlap $q_s=|\braket{s}{\psi}|^2$. The phase of $z_s$ records how the outcome sees the change. A real $z_s$ is a population change for that outcome, and an imaginary $z_s$ is a phase change. Then
\begin{equation}
\sum_sw_sz_s^2=0,\qquad \eta(u)\,\|u\|^2=\mathrm{Re}\sum_s\frac{w_sz_s^2}{q_s}.
\label{eq:denominator}
\end{equation}
The left identity holds for every 2-design, every state and every direction, because second moments make the population and phase responses cancel (Appendix~\ref{app:typical}). The right identity is the imbalance of Eq.~\eqref{eq:budget}, the same sum with each term divided by the overlap. When all overlaps are comparable, as on typical states, the division barely changes the sum and the cancellation survives. When a few outcomes carry large overlaps and many carry small ones, the division reweights the sum and destroys the cancellation. Superpositions of two measurement vectors, such as the encoded qubit, are of this kind.

\begin{result}{5 (typical states)}
For Haar-random pure states, the worst-direction fraction $\min_uI(u)/J(u)$ of complete-MUB measurements converges in probability to $1/2$ as $d\to\infty$. Its mean exceeds $(1-1/\sqrt2)/2\approx0.146$ in every dimension.
\end{result}

\begin{figure}[t]
\centering
\includegraphics[width=\columnwidth]{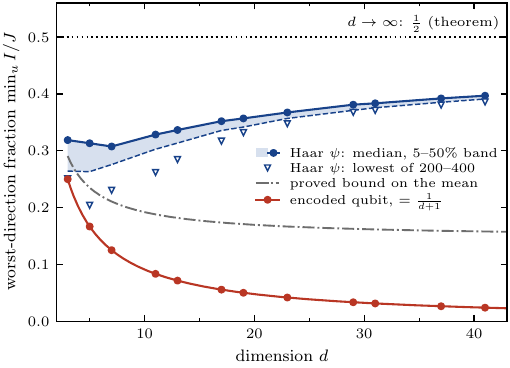}
\caption{\textbf{Typical versus structured states.} Worst-direction fraction $\min_uI/J$ under complete-MUB measurements at prime $d$. Blue symbols show Haar-random states, with the median (circles), the band down to the 5\% quantile (dashed edge), and the lowest of 200--400 samples (triangles). The mean approaches the asymptotic value $1/2$ (dotted, Result~5) as $1/2-0.65/\sqrt d$ (fit for $d\ge17$). The gray dash-dotted line is the proven lower bound on the mean, which tends to $(1-1/\sqrt2)/2$ (Appendix~\ref{app:typical}). The red line shows the encoded qubit at $1/(d+1)$, the exact worst case.}
\label{fig:typical}
\end{figure}

\noindent Numerically the mean approaches $1/2$ as $1/2-0.65/\sqrt d$ (Fig.~\ref{fig:typical}). The convergence comes from the phase factor $e^{-2i\arg\braket{s}{\psi}}$ that the imbalance attaches to every outcome. Its lowest harmonic is quadratic in the state, and the 2-design property cancels it, as in Eq.~\eqref{eq:denominator}. The higher harmonics fluctuate almost independently from outcome to outcome when the state is spread over many outcomes, so they average out (Appendix~\ref{app:typical}).

A benchmark on random states would therefore certify near-optimal performance and miss the loss entirely. At $d=41$ the median Haar-random state retains 0.40 of the QFI in its worst direction under complete-MUB measurements, while the encoded qubit retains $1/42\approx0.024$ (Fig.~\ref{fig:typical}). In practice, sensors prepare superpositions of two levels, and qudit processors store qubits in pairs of levels. Such states superpose two measurement vectors, the loss concentrates on them, and randomized measurements should be benchmarked on them.

Classical shadows show the same split between typical and worst cases. Every 2-design is optimal for typical observables, while complete MUBs and SICs need a factor of order $d$ more repetitions for the worst ones~\cite{Yang2026,Mao2025}. Section~\ref{sec:learning} turns this parallel into a quantitative correspondence.

\section{Measurements that retain a constant fraction}
\label{sec:breaking}

Since the loss comes from aligned blind spots, scattering them removes it. Random bases, magic gates and the Clifford circuits of qubit processors all scatter them.

\subsection{Random bases}
\label{sec:random}

Adding unbiased bases lowers the ceiling $1/K$ of Eq.~\eqref{eq:mub}, while adding independent random bases raises the retained fraction.

\begin{result}{6 (random bases)}
Mix $K\ge5$ independent Haar-random bases uniformly. There is a constant $c_K>0$, independent of $d$, such that in every sufficiently large dimension the measurement retains at least $c_K$ with probability at least $1-2^{-d}$. At least three bases are necessary.
\end{result}

\noindent The number five does not grow with $d$, while every 2-design built from bases needs at least $d+1$ of them (Appendix~\ref{app:extremal}). For the Fisher information, Result~6 puts a number, five, on the constant number of random bases that determine pure states stably~\cite{Voroninski2013}. With a number of random bases proportional to $d$, the retained fraction is at least $1/16$ (Appendix~\ref{app:random}). Two bases do not suffice, because a state that coincides with a vector of one basis leaves that basis with only second-order changes, and the other basis sees half of the directions.

Five bases suffice because independent bases rarely ignore the same signal. A random basis is nearly blind to a given signal only if the signal falls almost entirely into its blind spot. The blind spot is a random subspace of half the dimension, so a ratio $I/J$ of order $\varepsilon^2$ occurs with probability of order $(C\varepsilon)^{d-1}$ (Appendix~\ref{app:random}). For $K$ independent bases this happens together with probability of order $(C\varepsilon)^{K(d-1)}$. The signals to be protected are pairs of a state and a direction. Such a pair has about $4d$ real parameters, so at resolution $\varepsilon$ there are about $(1/\varepsilon)^{4d}$ of them. The bases protect every signal once
\begin{equation}
K(d-1)>4d ,
\label{eq:counting}
\end{equation}
which for large $d$ means $K\ge5$. Equation~\eqref{eq:counting} rests on independence alone.

\begin{figure}[t]
\centering
\includegraphics[width=\columnwidth]{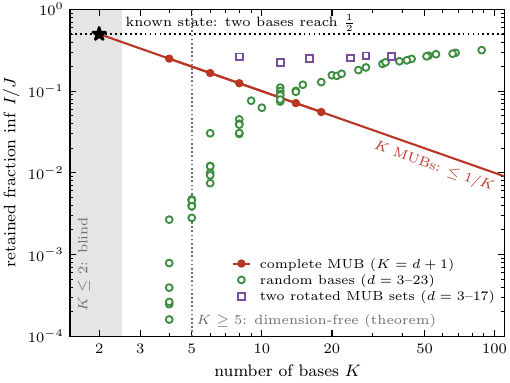}
\caption{\textbf{Alignment sets the cost.} Retained fraction versus the number of measurement bases $K$. In red, any $K$ mutually unbiased bases retain at most $1/K$ (line), and complete MUBs ($K=d+1$, filled circles) attain it. Green open circles show $K$ independent random bases for $d=3$--23 (adversarial search, minimum over draws). They fall on a common, dimension-independent curve that rises toward $1/2$. Five or more random bases provably retain a dimension-independent fraction (dotted vertical line). With $K\le2$ bases some direction of some state is unseen (gray). Purple open squares show two complete MUB sets related by a random unitary ($K=2(d+1)$, $d=3$--17). The star marks the Fisher-symmetric value $1/2$, which two bases reach for a known state.}
\label{fig:settings}
\end{figure}

Numerically, the retained fraction depends on the number of bases alone (Fig.~\ref{fig:settings}). Adversarial searches over states and directions give $c_K\approx0.004$, 0.01, 0.04 and 0.09 for $K=5$, 6, 8 and 12, flat from $d=5$ to $d=23$. The constants grow quickly with $K$, and about a dozen bases make the guarantee practical. Twelve random bases cost at most about eleven times the quantum limit in every dimension tested, overtake complete MUBs near $d\approx11$, and gain linearly beyond. For twelve bases at $d=3$, an interval branch-and-bound certificate places the true worst case within 10\% of the search value (Appendix~\ref{app:numerics}). Two complete MUB sets related by a random unitary also scatter the blind spots and retain about one quarter of the QFI in every dimension tested ($d\le17$).

\subsection{Relation to shadow estimation}
\label{sec:learning}

Quantum learning gives a second view of why scattering helps. A classical shadow estimates an observable $O$ from randomized measurement records, and its shadow norm $\|O\|_{\mathrm{sh}}$ sets the variance of this estimate in the worst case~\cite{Huang2020shadows}. Let $\kappa$ be the largest ratio $\|O\|_{\mathrm{sh}}^2/\|O\|_2^2$ over traceless observables. It is at most 3 for 3-designs and at most $d+1$ for 2-designs~\cite{Huang2020shadows,Yang2026}. Building on the Cauchy--Schwarz argument of Refs.~\cite{Du2026,Cho2026}, we turn this learning cost into a sensing guarantee.

\begin{result}{7 (correspondence between sensing and learning)}
Every randomized measurement that forms a 2-design satisfies
\begin{equation}
I\succeq\frac{J}{2\kappa}.
\label{eq:bridge}
\end{equation}
Random Clifford measurements of $n$ qudits of odd prime dimension $p$ therefore retain between $1/(4p-2)$ and $1/(p+1)$ for every $n$, and adding a single magic gate retains at least $1/15$.
\end{result}

\noindent The proof uses the shadow estimate of the operator $\partial\rho$, which carries the signal, as a test function in the Cauchy--Schwarz bound on Fisher information (Appendix~\ref{app:learning}). Physically, sensing a parameter with a randomized measurement costs no more than learning the signal operator $\partial\rho$ with the same measurement. For 2-designs Eq.~\eqref{eq:bridge} reproduces the floor of Result~4(b). For qudit Clifford measurements the shadow norm gives $\kappa\le2p-1$~\cite{Mao2025}, and the encoded qubit of Sec.~\ref{sec:mub} gives the upper value $1/(p+1)$. Neither bound depends on the number of qudits. A single magic gate bounds $\kappa$ by $15/2$~\cite{Zhu2024magic}. Numerically the magic-gate measurement retains between 0.33 and 0.44 in its worst case for $d\le11$.

Sensing can also be much cheaper than learning. Near-optimal sensing of every pure state needs $O(1)$ random bases (Result~6), whereas worst-case-optimal shadow estimation needs of order $d^2$ bases~\cite{Yang2026}.

\begin{figure*}[t]
\centering
\includegraphics[width=\textwidth]{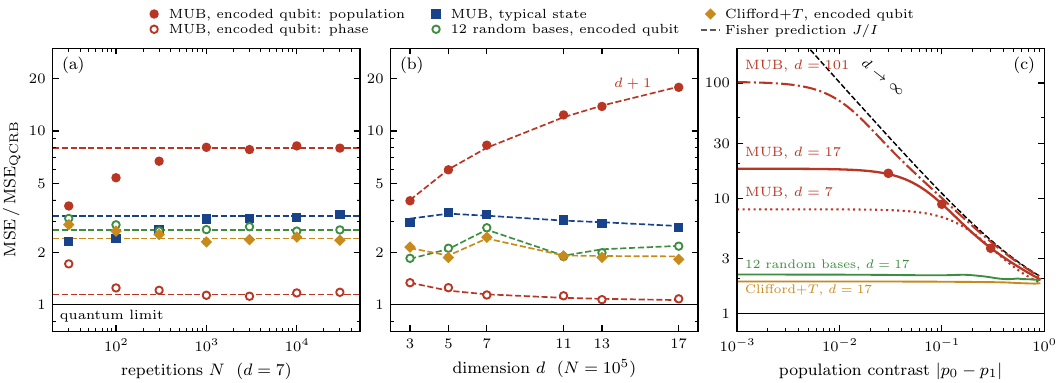}
\caption{\textbf{Fisher information predicts the cost in repetitions.} Mean-squared error of maximum-likelihood estimation, in units of the quantum Cram\'er--Rao bound $1/(NJ)$, from 4000 simulated experiments per point. Error bars show one standard error. Dashed lines give the Fisher prediction $J/I$. Unless noted, the state is a qubit encoded in two levels of a qudit, $\psi=(\ket0+\ket1)/\sqrt2$, and the parameter is its population (filled red) or its phase (open red). (a)~Convergence with the number of repetitions at $d=7$. Complete-MUB measurements need eight times more repetitions than the quantum limit for the population and 1.14 times for the phase. On the same state, twelve random bases and Clifford circuits with one $T$ gate cost about 2--3, as does a typical state measured with MUBs along its worst direction. At small $N$ the population estimates are biased toward equal populations, which lowers the error. (b)~At $N=10^5$, the population cost of MUB measurements grows as $d+1$, from 4 at $d=3$ to 18 at $d=17$, while the other measurements stay near 2--3. (c)~Dependence on the operating point. The MUB population cost is $d+1$ at equal populations and approaches $(1+t)/t$ (dashed, $d\to\infty$) once the population contrast $t=|p_0-p_1|$ exceeds about $1/d$. Points are simulations at $d=17$, $N=10^5$. Random-basis and Clifford$+T$ measurements keep the same cost at every operating point.}
\label{fig:finite}
\end{figure*}

\subsection{Random Clifford measurements of qubits}
\label{sec:clifford}

\begin{result}{8 (random Clifford measurements of qubits)}
Random Clifford measurements of $n$ qubits retain exactly $c=1/3$ for every $n$. At every stabilizer state the Fisher information is isotropic, $I=J/3$ in every direction. At nearby states the worst direction retains $1/3$ and its conjugate $2/3$.
\end{result}

\noindent Random Clifford circuits are the standard randomizer of qubit processors, and however many qubits are added, their worst case stays that of a single qubit measured along its three Pauli axes. Result~8 sharpens the general 3-design guarantee, which tends to $1/4$ for many qubits~\cite{Zhou2026}, into an exact value for the most widely used randomized measurement. At stabilizer states the value follows from a $q$-binomial count of the stabilizer states orthogonal to a given one (Appendix~\ref{app:clifford}). Stabilizer states make many outcomes dark, so their budget is not full, $I(u)+I(iu)=2J(u)/3$. At nearby states the budget is full again and splits into $1/3$ and $2/3$. Away from stabilizer states the proof requires a recursion over Pauli observables.

The single-qubit value survives at any size because drawing a random stabilizer basis is equivalent to measuring a random Pauli observable first and then continuing in the subspace it selects. The chain rule of Fisher information turns this into a recursion over the number of qubits. The recursion closes if a random Pauli measurement is at most twice as sensitive to the conjugate change $iu$ as to $u$. This bound comes from an identity for pure states,
\begin{equation}
\sum_{Q:\{P,Q\}=0}\langle Q\rangle^2=\frac D2\bigl(1-\langle P\rangle^2\bigr),
\label{eq:pauli}
\end{equation}
where the sum runs over the Pauli operators $Q$ that anticommute with a Pauli operator $P$, and $D=2^n$. By Eq.~\eqref{eq:pauli}, the more definite $P$ is, the less weight its anticommuting partners can carry. Two nearly definite Pauli observables must therefore commute, which keeps the alignment of blind spots at the single-qubit level.

\section{Design rules and examples}
\label{sec:design}

The design rule is to randomize so that blind spots scatter. On qubits, random Clifford measurements are safe, and Result~8 fixes their cost at three times the quantum limit in the worst direction. On qudits, complete-MUB and random Clifford measurements cost a factor $d+1$ for population-type parameters. One magic gate restores a dimension-independent fraction, and so do random bases, of which about a dozen beat complete MUBs from $d\approx11$ on. For photonic modes, programmable light converters~\cite{Lib2025photonic} and complex media~\cite{Goel2024} already implement Haar-random bases. When a rough estimate of the state is available, two bases tailored to it approach the optimum (Result~2). Any benchmark of a randomized measurement should include structured states, where the loss concentrates.

For a qubit encoded in two levels of a trapped ion or in two modes of a photon, complete-MUB or random Clifford measurements retain $d/(d+1)$ of the QFI for phase-type parameters. Population-type parameters retain $1/(d+1)$ at equal populations (Sec.~\ref{sec:mub}). On qubit processors, Result~8 makes the error budget of Hamiltonian estimation with random stabilizer input states and random Clifford measurements exact~\cite{Zhou2026}. The Fisher information equals $J/3$ in every direction, so every parameter costs exactly three times the quantum limit.

Finite-sample simulations confirm that the Fisher information predicts the actual cost of an experiment (Fig.~\ref{fig:finite}). We simulate maximum-likelihood estimation from up to $10^5$ repetitions and compare its mean-squared error with the quantum Cram\'er--Rao bound $1/(NJ)$. For the population of a qubit encoded in a $d=7$ qudit, complete-MUB measurements settle at eight times the quantum limit once $N\gtrsim10^3$, while the conjugate phase costs 1.14 times the limit [Fig.~\ref{fig:finite}(a)]. On the same state, twelve random bases cost 2.7, about a quarter of their worst-case cost (Sec.~\ref{sec:random}), and Clifford circuits with a single magic gate, here a $T$ gate, cost 2.4. Across dimensions the MUB population cost tracks $d+1$, while the other measurements stay near 2--3 [Fig.~\ref{fig:finite}(b)]. The Cram\'er--Rao bound covers every unbiased estimator and maximum likelihood attains it asymptotically, so this gap is set by the measurement itself. In large dimensions the simulated cost reaches the Fisher prediction only once $N\gtrsim d^3$, because a few interference outcomes are nearly dark and the second-order change of the visibility alters their tiny probabilities by a large relative amount. These nearly dark outcomes pull finite-$N$ estimates toward equal populations (Appendix~\ref{app:numerics}), an effect common in finite-sample metrology~\cite{Meyer2025}.

The blindness is sharpest at equal populations. Moving the operating point to a population contrast $t=|P|$ lets the darkest outcomes of the interference bases regain sensitivity, and the cost falls toward $(1+t)/t$ once $t\gtrsim1/d$ [Fig.~\ref{fig:finite}(c)]. A sensor operated near equal populations therefore pays about $\min[d+1,(1+t)/t]$, close to elevenfold at $t=0.1$ in large dimensions. Random-basis and magic-gate measurements keep the same cost at every operating point.

The blind-spot geometry carries over to phase retrieval, the recovery of a signal from intensities. Phase retrieval measures the intensities $|\braket{s}{x}|^2$ of a vector $x$ against a frame of test vectors and reconstructs $x$ up to a global phase~\cite{Shechtman2015,Dong2023}. The worst Fisher ratio of the frame is the local stability constant of this reconstruction~\cite{Balan2016}, which measures how strongly the intensities react to the worst small change of the signal. Frames built from MUBs or SICs are therefore locally unstable, with a stability constant of order $1/d$ at signals that superpose two frame vectors. Frames of random bases keep it independent of the dimension, and five bases suffice (Result~6).

\section{Discussion}
\label{sec:discussion}

The cost of a randomized measurement is set by how the blind spots of its bases are arranged, and second moments leave this arrangement undetermined. Complementarity aligns the blind spots of unbiased bases, because a population change for one basis is a phase change for every other. The symmetry that makes complete MUBs optimal for tomography thus gives them the largest loss among 2-designs built from bases, a factor of the dimension plus one. This factor has the order of the overhead that learning finds for second-order designs, and in sensing the loss concentrates on structured states, which averages over random states miss. Scattering the blind spots removes it, and, as in phase retrieval, a few unstructured bases suffice. For an unknown state of many levels, a dozen random bases therefore outperform a complete set of unbiased bases from about eleven levels on, and the gain grows linearly with the dimension.

Superpositions of two measurement vectors, on which the loss concentrates, also locate it in other structured randomizations. For Clifford circuits on many qudits of odd prime dimension, a qubit encoded in one qudit costs the local dimension plus one however many qudits are added, and numerics on two qudits find no worse state. We expect this value to be exact, and Appendix~\ref{app:clifford} reduces its proof to a single inequality on Pauli lines, the qudit form of the bound that closes the qubit recursion of Sec.~\ref{sec:clifford}. For local randomized measurements, the most common randomizer in experiments~\cite{Brydges2019,Elben2023toolbox}, the GHZ state of $n$ qubits superposes two measurement vectors, $\ket{0\cdots0}$ and $\ket{1\cdots1}$. Every local Pauli setting that measures some qubit along $Z$ tells these two vectors apart and is blind to their relative phase. Only the settings that measure every qubit along $X$ or $Y$ see this phase. They form a fraction $(2/3)^n$, so random local Pauli measurements retain at most $(2/3)^n$ of the QFI of a GHZ sensor. Its QFI grows as $n^2$, while the Fisher information of this readout falls exponentially with $n$. How many entangling layers scatter the blind spots of local measurements is open, and circuits of logarithmic depth already form approximate designs~\cite{Schuster2025lowdepth}.

Scattering, in turn, needs little randomness. Three random bases are necessary and five suffice in large dimensions, and whether three or four suffice is open. Structured sets scatter as well, since two complete MUB sets related by one random unitary retain about a quarter of the QFI in every dimension tested. This makes $2(d+1)$ bases with a dimension-independent guarantee a concrete target for a proof.

On mixed states with every parameter unknown, no single-copy measurement can in general retain a dimension-independent fraction of the QFI~\cite{Gill2000,Zhou2026}, so Ref.~\cite{Zhou2026} extended the 2-design question to states of any rank, using a weaker sense of near-optimality. Pure states are the rank-one members of this family, and on them the weaker sense coincides with a dimension-independent retained fraction~\cite{Zhou2026}. The losses of complete MUBs and SICs therefore answer the extended question in the negative as well (Appendix~\ref{app:mub}).

\begin{acknowledgments}
This work is supported by the Fundamental and Interdisciplinary Disciplines Breakthrough Plan of the Ministry of Education of China under Grant No.~JYB2025XDXM606 and the National Natural Science Foundation of China under Grant No.~42388101.
\end{acknowledgments}

\section*{Data availability}
No new data were created or analyzed in this study.

\section*{Appendix}
\appendix
\section{Notation, conventions and summary of results}
\label{app:notation}

This appendix fixes the notation used in all proofs. The symbols of the main text keep their meaning, and $\psi\in\mathbb C^d$ always denotes a unit vector.

\paragraph{States and directions.}
A pure-state model $\theta\mapsto\psi_\theta$ with $\theta\in\mathbb R^m$ has, at a given point, the tangent vectors $u_j=(1-\ket\psi\bra\psi)\partial_j\psi$, where the projector removes the global-phase part. A real vector $v\in\mathbb R^m$ selects the direction $u_v=\sum_jv_ju_j$, and the QFI matrix satisfies $v^{\mathsf T}Jv=J(u_v)=4\|u_v\|^2$. The Fisher information matrix of a measurement satisfies $v^{\mathsf T}Iv=I(u_v)$. An inequality $I(u)\ge c\,J(u)$ for all $u\perp\psi$ is therefore the same as the matrix inequality $I\succeq cJ$ for every pure-state model.

\paragraph{Measurements.}
All measurements are rank-one positive operator-valued measures (POVMs),
\begin{equation}
M=\{w_s\ket s\bra s\},\qquad \sum_sw_s\ket s\bra s=\openone ,
\label{eq:A:povm}
\end{equation}
with weights $w_s>0$ and unit vectors $\ket s$. Every POVM refines into a rank-one POVM by splitting its elements, and refinement does not decrease Fisher information.
A randomized measurement over orthonormal bases $\{\ket{b_{a,x}}\}_x$, chosen with probabilities $\lambda_a$, is the POVM with outcomes $s=(a,x)$, weights $w_s=\lambda_a$ and vectors $\ket s=\ket{b_{a,x}}$, so the basis label is part of the outcome. We write
\begin{equation}
\begin{aligned}
q_s&=|\braket{s}{\psi}|^2, & p_s&=w_sq_s,\\
z_s&=\braket{\psi}{s}\braket{s}{u}, & \partial_up_s&=2w_s\,\mathrm{Re}\,z_s .
\end{aligned}
\label{eq:A:notation}
\end{equation}

\paragraph{Fisher information.}
Following Ref.~\cite{Zhou2026}, outcomes with $q_s=0$ are removed,
\begin{equation}
I(u)=\sum_{s:\,q_s>0}\frac{(\partial_up_s)^2}{p_s}=4\sum_{s:\,q_s>0}\frac{w_s(\mathrm{Re}\,z_s)^2}{q_s}.
\label{eq:A:fisher}
\end{equation}
Outcomes with $q_s=0$ have $\partial_up_s=0$, because the nonnegative function $q_s$ has a minimum at $\psi$. We call an outcome dark at $\psi$ when $q_s=0$, and we call $\psi$ regular for $M$ when no outcome is dark. Along a smooth path, a dark outcome with $\braket{s}{u}\ne0$ becomes possible at second order and adds $4w_s|\braket{s}{u}|^2$ to the Fisher information of the path. We call the Fisher information of the path the pathwise Fisher information, as opposed to the pointwise value~\eqref{eq:A:fisher}. Appendix~\ref{app:numerics} uses the pathwise value to compare with simulations. Lemma~B3 shows that the retained fraction of Eq.~\eqref{eq:retained} equals the infimum over points $(\psi,u)$ where no dark outcome responds to $u$. The pathwise and pointwise values agree at these points, and the pathwise value is never smaller, so both give the same worst-case values.

\paragraph{Designs.}
A rank-one POVM is a complex projective $t$-design when
\begin{equation}
\sum_sw_s\bigl(\ket s\bra s\bigr)^{\otimes t}=d\binom{d+t-1}{t}^{-1}P^{(t)}_{\mathrm{sym}},
\label{eq:A:design}
\end{equation}
where $P^{(t)}_{\mathrm{sym}}$ projects onto the symmetric subspace of $(\mathbb C^d)^{\otimes t}$. For $t=2$ the condition reads $\sum_sw_s(\ket s\bra s)^{\otimes2}=2P_{\mathrm{sym}}/(d+1)$ with $P_{\mathrm{sym}}=(\openone+\mathbb F)/2$ and $\mathbb F$ the swap. A 2-design built from orthonormal bases is a randomized measurement over bases that satisfies Eq.~\eqref{eq:A:design} for $t=2$. Complete MUBs with $\lambda_a=1/(d+1)$ are the standard example~\cite{Klappenecker2005}. A SIC is the 2-design with $d^2$ outcomes, $w_s=1/d$ and $|\braket{s}{t}|^2=1/(d+1)$ for $s\ne t$~\cite{Renes2004}.

\paragraph{Clifford measurements.}
For a prime $p$ and $n$ qudits we write $D=p^n$. Stabilizer states are the joint eigenstates of maximal commuting subgroups of the Pauli (Weyl) group~\cite{Gottesman1998,Dehaene2003,Hostens2005}, and there are $N_{n,p}=D\prod_{j=1}^n(p^j+1)$ of them. A random Clifford measurement applies a uniformly random Clifford unitary and then measures in the computational basis. The Clifford group acts transitively on stabilizer states, so the measurement is the rank-one POVM of all stabilizer states with equal weights $w=D/N_{n,p}$.

\paragraph{Summary of results.}
Table~\ref{tab:A1} collects the retained fractions proved or computed in this work, grouped by the mechanism of Sec.~\ref{sec:half}. The figures of the main text display the same values.

\begin{table*}[t]
\caption{\textbf{Retained fractions of randomized measurements on pure states.} Values marked ``$=$'' are exact, ``$\ge$'' and ``$\le$'' are proven bounds, and ``$\approx$'' are numerical. ``Aligned'' measurements leave a common blind spot for some state and direction, and ``scattered'' measurements do not.}
\label{tab:A1}
\begin{ruledtabular}
\begin{tabular}{p{5.2cm}p{6.6cm}p{2.4cm}p{2.8cm}}
Measurement & Retained fraction $c$ & Mechanism & Source \\
\hline
Known state, two bases & every allocation with $I(u)+I(iu)=J(u)$, including the Fisher-symmetric value $1/2$ & known state & Thm.~2 \\
Haar-random basis & $=1/2$ in every direction & scattered & Refs.~\cite{Hayashi1998,Lu2025} \\
Any 3-design & $\ge(d+2)/[4(d+1)]$ & scattered & Ref.~\cite{Zhou2026}, App.~\ref{app:learning} \\
Random Clifford, $n$ qubits & $=1/3$ for every $n$, isotropic at stabilizer states & scattered & Thm.~8 \\
Clifford circuits with one magic gate & $\ge1/15$, and $\approx0.33$--$0.44$ for $d\le11$ & scattered & Cor.~G2, App.~\ref{app:numerics} \\
$K\ge5$ random bases & $\ge c_K>0$ independent of $d$, and $c_{12}\approx0.09$ & scattered & Thm.~6, App.~\ref{app:numerics} \\
$K\ge570\,d$ random bases & $\ge1/16$ & scattered & Prop.~F5 \\
Two MUB sets related by a random unitary & $\approx0.22$--$0.27$ for $d\le17$ & scattered & App.~\ref{app:numerics} \\
Complete MUBs and random Clifford measurements of one qudit of odd prime $d$ & $=1/(d+1)$ & aligned & Thms.~3, 4 \\
Any $K$ MUBs, any weights & $\le\min_a\lambda_a\le1/K$ & aligned & Thm.~3 \\
Any 2-design built from bases & $\ge1/(d+1)$ & --- & Thm.~4 \\
SIC & $=(1-1/\sqrt{d+1})/d$ & aligned & Thm.~4 \\
Any 2-design & $\ge1/[2(d+1)]$ & --- & Thm.~4 \\
Random Clifford, $n$ qudits of odd prime $p$ & $\in[1/(4p-2),\,1/(p+1)]$ for every $n$ & aligned & Thm.~3, Cor.~G1 \\
Complete MUBs, Haar-random state & $\to1/2$ in probability, with mean $\ge(1-1/\sqrt2)/2$ & scattered (state) & Thm.~5 \\
\end{tabular}
\end{ruledtabular}
\end{table*}

\section{Information budget and the known-state region}
\label{app:budget}

This appendix proves Results~1 and~2. It also gives the matrix form of the imbalance used in later appendices and shows that a dark outcome is an imbalance seen from a nearby state.

\subsection{The budget identity}

\begin{theorem}{1 (information budget)}
Let $M$ be a rank-one POVM, $\psi$ a pure state and $u\perp\psi$ nonzero. Define the leaked fraction and the imbalance
\begin{equation}
\begin{aligned}
\ell(u)&=\frac{1}{\|u\|^2}\sum_{s:\,q_s=0}w_s|\braket{s}{u}|^2,\\
\eta(u)&=\frac{1}{\|u\|^2}\,\mathrm{Re}\sum_{s:\,q_s>0}\frac{w_sz_s^2}{q_s}.
\end{aligned}
\label{eq:B:defs}
\end{equation}
Then
\begin{equation}
\frac{I(u)}{J(u)}=\frac12\bigl[1-\ell(u)+\eta(u)\bigr],
\label{eq:B:budget}
\end{equation}
with $\ell(iu)=\ell(u)$ and $\eta(iu)=-\eta(u)$. Therefore $I(u)+I(iu)=J(u)\,[1-\ell(u)]$. At regular points $\ell\equiv0$ and $I(u)+I(iu)=J(u)$.
\end{theorem}

\begin{proof}
Write $(\mathrm{Re}\,z)^2=\frac12(|z|^2+\mathrm{Re}\,z^2)$ in Eq.~\eqref{eq:A:fisher}. Since $|z_s|^2/q_s=|\braket{s}{u}|^2$, completeness~\eqref{eq:A:povm} gives
\begin{equation*}
\sum_{s:\,q_s>0}w_s\frac{|z_s|^2}{q_s}=\|u\|^2-\sum_{s:\,q_s=0}w_s|\braket{s}{u}|^2=\|u\|^2\bigl[1-\ell(u)\bigr].
\end{equation*}
Dividing $I(u)=2\|u\|^2[1-\ell(u)+\eta(u)]$ by $J(u)=4\|u\|^2$ gives Eq.~\eqref{eq:B:budget}. Replacing $u$ by $iu$ leaves $|\braket{s}{u}|$ unchanged and multiplies every $z_s^2$ by $-1$.
\end{proof}

At regular points the identity $I(u)+I(iu)=J(u)$ is the exact pure-state form of the Gill--Massar bound~\cite{Gill2000}, written per direction as in Ref.~\cite{Lu2021}. The symmetric point $\eta\equiv0$ is the Fisher-symmetric measurement of Ref.~\cite{Li2016}.

For one orthonormal basis, Theorem~1 reduces to Malus's law of Eq.~\eqref{eq:malus}. Let $\{\ket x\}$ be the basis and $\psi$ a state with all $q_x>0$, and define the population subspace
\begin{equation}
L(\psi)=\Bigl\{v\perp\psi:\ \frac{\braket{x}{v}}{\braket{x}{\psi}}\in\mathbb R\ \text{for all }x\Bigr\}.
\label{eq:B:population}
\end{equation}

\begin{corollary}{B1 (Malus's law)}
$L(\psi)$ and $iL(\psi)$ have real dimension $d-1$ each and are orthogonal for the real inner product $\mathrm{Re}\langle\cdot,\cdot\rangle$, so the tangent space is $L(\psi)\oplus iL(\psi)$. If $\vartheta$ is the angle between $u$ and $L(\psi)$, the basis gives $I_B(u)/J(u)=\cos^2\vartheta$ and $\eta_B(u)=\cos2\vartheta$. A randomized measurement over bases at a regular point has $\eta(u)=\sum_a\lambda_a\cos2\vartheta_a$. Malus's law for one basis and this formula are Eqs.~\eqref{eq:malus}--\eqref{eq:budget} of the main text.
\end{corollary}

\begin{proof}
Write $\braket{x}{\psi}=r_xe^{i\phi_x}$ and $\braket{x}{u}=e^{i\phi_x}(\alpha_x+i\beta_x)$. The condition $u\perp\psi$ reads $\sum_xr_x\alpha_x=\sum_xr_x\beta_x=0$, so the vectors $u_\alpha$ and $u_\beta$ with components $e^{i\phi_x}\alpha_x$ and $ie^{i\phi_x}\beta_x$ are both orthogonal to $\psi$. Each of $L(\psi)$ and $iL(\psi)$ is thus parametrized by a vector of $\mathbb R^d$ under one linear constraint and has real dimension $d-1$. By construction $u_\alpha\in L(\psi)$, $u_\beta\in iL(\psi)$, $u=u_\alpha+u_\beta$ and $\mathrm{Re}\langle u_\alpha,u_\beta\rangle=\mathrm{Re}\sum_x\alpha_x(i\beta_x)=0$. Hence $\|u_\alpha\|^2=\|\alpha\|^2=\|u\|^2\cos^2\vartheta$ and $\|\beta\|^2=\|u\|^2\sin^2\vartheta$. Finally $z_x=r_x(\alpha_x+i\beta_x)$, so $\eta_B(u)\|u\|^2=\sum_x(\alpha_x^2-\beta_x^2)$ and Eq.~\eqref{eq:B:budget} gives $I_B/J=\cos^2\vartheta$. The Fisher information of a randomized measurement is the $\lambda_a$-weighted sum over bases.
\end{proof}

\subsection{Matrix form of the imbalance}

The worst direction at a given state minimizes $I(u)/J(u)$, a ratio of two quadratic forms in $u$. To write them as matrices, fix an isometry $T:\mathbb C^{d-1}\to\psi^\perp$, so that $T^\dagger T=\openone$ and $TT^\dagger=\openone-\ket\psi\bra\psi$. Write $u=Tz$ and $v_s=T^\dagger\ket s$, and define
\begin{equation}
G_\psi=\sum_{s:\,q_s>0}w_s\,v_sv_s^\dagger,\quad
B_\psi=\sum_{s:\,q_s>0}w_s\,e^{-2i\arg\braket{s}{\psi}}\,\bar v_s v_s^\dagger .
\label{eq:B:takagi}
\end{equation}
$G_\psi$ is Hermitian with $0\preceq G_\psi\preceq\openone$, and $B_\psi=B_\psi^{\mathsf T}$ is complex symmetric. Since $v_s^\dagger z=\braket{s}{u}$ and $e^{-2i\arg\braket{s}{\psi}}\braket{s}{u}^2=z_s^2/q_s$,
\begin{equation}
\frac{I(u)}{J(u)}=\frac{z^\dagger G_\psi z+\mathrm{Re}\,(z^{\mathsf T}B_\psi z)}{2\,z^\dagger z}.
\label{eq:B:quadratic}
\end{equation}
At regular points $G_\psi=\openone$. A change of frame $T\mapsto TV$ with unitary $V$ maps $B_\psi\mapsto V^{\mathsf T}B_\psi V$, which leaves its Takagi values unchanged~\cite{HornJohnson2012}.

\begin{proposition}{B2 (spectrum)}
At a regular point the normalized Fisher matrix $J^{-1/2}IJ^{-1/2}$, a real symmetric form on the $2(d-1)$-dimensional real tangent space, has eigenvalues $(1\pm\sigma_k)/2$, $k=1,\dots,d-1$, where $\sigma_k\in[0,1]$ are the Takagi values of $B_\psi$. In particular $\min_uI(u)/J(u)=(1-\sigma_{\max})/2$. For one orthonormal basis $\sigma_k=1$ for all $k$, and the normalized Fisher matrix is the projector onto $L(\psi)$.
\end{proposition}

\begin{proof}
Take a Takagi factorization $B_\psi=U\Sigma U^{\mathsf T}$ with unitary $U$ and $\Sigma=\mathrm{diag}(\sigma_k)$~\cite{HornJohnson2012}. With $z=\bar Uy$ one has $z^\dagger z=y^\dagger y$ and $z^{\mathsf T}B_\psi z=y^{\mathsf T}\Sigma y$. Writing $y_k=x_k+it_k$ gives $\mathrm{Re}\,y^{\mathsf T}\Sigma y=\sum_k\sigma_k(x_k^2-t_k^2)$, so Eq.~\eqref{eq:B:quadratic} is diagonal with eigenvalues $(1\pm\sigma_k)/2$. Positivity of the Fisher matrix gives $\sigma_k\le1$. For one basis, Corollary~B1 shows that the normalized Fisher matrix is the projector onto $L(\psi)$, whose eigenvalues are $0$ and $1$.
\end{proof}

\subsection{Dark outcomes}

Dark outcomes that respond to $u$ make the budget fall short, $\ell(u)>0$, but they do not lower the retained fraction.

\begin{lemma}{B3 (approach without leakage)}
Let $\psi_0$ be a pure state and $w\perp\psi_0$ a unit vector. There are points $(\psi_\varepsilon,w_\varepsilon)\to(\psi_0,w)$ with $\ell_\varepsilon(w_\varepsilon)=0$ and
\begin{equation*}
\lim_{\varepsilon\to0}\frac{I(w_\varepsilon)}{J(w_\varepsilon)}=\frac{I(w)}{J(w)}\Big|_{\psi_0}.
\end{equation*}
Consequently the retained fraction~\eqref{eq:retained} equals the infimum over points with $\ell=0$, where $I(u)+I(iu)=J(u)$ and every loss is an imbalance.
\end{lemma}

\begin{proof}
Let $v=-iw$, $\psi_\varepsilon=\cos\varepsilon\,\psi_0+\sin\varepsilon\,v$, $u_\varepsilon=-\sin\varepsilon\,\psi_0+\cos\varepsilon\,v$ and $w_\varepsilon=iu_\varepsilon$, so that $w_\varepsilon\perp\psi_\varepsilon$ and $w_\varepsilon\to iv=w$. For an outcome that is dark at $\psi_0$, $\braket{s}{\psi_\varepsilon}=\sin\varepsilon\braket{s}{v}$ and $\braket{s}{w_\varepsilon}=i\cos\varepsilon\braket{s}{v}$. If $\braket{s}{v}=0$, the outcome stays dark and does not respond to $w_\varepsilon$, so it contributes neither to $\ell_\varepsilon$ nor to $\eta_\varepsilon$. If $\braket{s}{v}\ne0$, the outcome is possible for $\varepsilon\ne0$, with $z_s=i\sin\varepsilon\cos\varepsilon|\braket{s}{v}|^2$ and
\begin{equation*}
\frac{z_s^2}{q_s}=-\cos^2\varepsilon\,|\braket{s}{v}|^2\ \longrightarrow\ -|\braket{s}{w}|^2 .
\end{equation*}
For outcomes possible at $\psi_0$, $z_s^2/q_s$ is continuous in $\varepsilon$ and tends to $(\braket{\psi_0}{s}\braket{s}{w})^2/q_s(\psi_0)$. Hence $\ell_\varepsilon(w_\varepsilon)=0$ and $\eta_\varepsilon(w_\varepsilon)\to\eta_0(w)-\ell_0(w)$, and Eq.~\eqref{eq:B:budget} gives the limit $\frac12[1-\ell_0(w)+\eta_0(w)]$, which is the value at $\psi_0$.
\end{proof}

The same computation in the direction of approach shows that near $\psi_0$ the dark outcomes add $\ell_0$ to the imbalance of $u_\varepsilon$ and remove it from the conjugate direction. A dark outcome is therefore an imbalance seen from a nearby state.

\subsection{The known-state region}

Theorem~2 below characterizes the normalized Fisher matrices that single-copy measurements reach at a known state. To state it, identify the tangent space with $\mathbb R^{2(d-1)}$ through $\hat z=(\mathrm{Re}\,z,\mathrm{Im}\,z)$, and let $\mathcal J=\bigl(\begin{smallmatrix}0&-\openone\\ \openone&0\end{smallmatrix}\bigr)$ represent multiplication by $i$. A Hermitian matrix $G$ and a complex symmetric matrix $B$ have the real forms
\begin{equation}
\hat G=\begin{pmatrix}\mathrm{Re}\,G&-\mathrm{Im}\,G\\ \mathrm{Im}\,G&\mathrm{Re}\,G\end{pmatrix},\qquad
H_B=\begin{pmatrix}\mathrm{Re}\,B&-\mathrm{Im}\,B\\ -\mathrm{Im}\,B&-\mathrm{Re}\,B\end{pmatrix},
\label{eq:B:realforms}
\end{equation}
with $z^\dagger Gz=\hat z^{\mathsf T}\hat G\hat z$ and $\mathrm{Re}\,z^{\mathsf T}Bz=\hat z^{\mathsf T}H_B\hat z$. The real forms satisfy $\mathcal J^{\mathsf T}\hat G\mathcal J=\hat G$ and $\mathcal J^{\mathsf T}H_B\mathcal J=-H_B$, and every real symmetric $X$ splits uniquely as $X=\frac12(X+\mathcal J^{\mathsf T}X\mathcal J)+\frac12(X-\mathcal J^{\mathsf T}X\mathcal J)$ into a part of the form $\hat G$ and a part of the form $H_B$. By Eq.~\eqref{eq:B:quadratic}, the normalized Fisher matrix of a rank-one POVM is
\begin{equation}
F=\tfrac12\bigl(\hat G_\psi+H_{B_\psi}\bigr).
\label{eq:B:F}
\end{equation}

\begin{theorem}{2 (known state)}
Fix a pure state $\psi$.
(a)~A real symmetric $F$ is the normalized Fisher matrix of some single-copy measurement at $\psi$ if and only if
\begin{equation}
F\succeq0,\qquad F+\mathcal J^{\mathsf T}F\mathcal J\preceq\openone .
\label{eq:B:region}
\end{equation}
(b)~The boundary $F+\mathcal J^{\mathsf T}F\mathcal J=\openone$, that is, the allocations with $I(u)+I(iu)=J(u)$ for all $u$, consists of $F=\frac12(\openone+H_B)$ with $B$ any complex symmetric contraction. Every boundary point is attained by an equal-weight mixture of two orthonormal bases, both unbiased to $\psi$.
(c)~One orthonormal basis attains exactly the boundary points with $B$ unitary, which are the projectors onto $(d-1)$-dimensional subspaces $L$ with $L\perp iL$.
(d)~For every pure-state model and every positive weight matrix $W$, the Holevo bound is attained by a random choice between two orthonormal bases of the system.
\end{theorem}

\begin{proof}
\emph{Necessity in (a).} For a rank-one POVM, Eq.~\eqref{eq:B:F} gives $F+\mathcal J^{\mathsf T}F\mathcal J=\hat G_\psi\preceq\openone$, because $G_\psi\preceq\sum_sw_sv_sv_s^\dagger=\openone$. A general POVM with Fisher matrix $F$ refines into a rank-one POVM $M'$ with Fisher matrix $F'\succeq F$, and $X\mapsto X+\mathcal J^{\mathsf T}X\mathcal J$ preserves the order.

\emph{Sufficiency in (a).} Let $F$ satisfy~\eqref{eq:B:region}. Then $\hat G:=F+\mathcal J^{\mathsf T}F\mathcal J$ commutes with $\mathcal J$ and is the real form of a Hermitian $G$ with $0\preceq G\preceq\openone$, and $F-\mathcal J^{\mathsf T}F\mathcal J$ anticommutes with $\mathcal J$ and equals $H_B$ for a complex symmetric $B$. Factor $F=LL^{\mathsf T}$ with $L\in\mathbb R^{2(d-1)\times r}$, $r=\mathrm{rank}\,F$. Let $N=r+1$, $\rho=(1,\dots,1)/\sqrt N\in\mathbb R^N$, and let $O\in\mathbb R^{r\times N}$ have orthonormal rows orthogonal to $\rho$. Put $\hat W=LO$, split it into its upper and lower halves, and define $V=\hat W_{\rm up}-i\hat W_{\rm low}\in\mathbb C^{(d-1)\times N}$. A direct expansion shows $\hat W\hat W^{\mathsf T}=\frac12(\hat G'+H_{B'})$ with $G'=\overline{VV^\dagger}$ and $B'=VV^{\mathsf T}$. Since $\hat W\hat W^{\mathsf T}=F$ and $F$ splits uniquely as $\frac12(\hat G+H_B)$, $G'=G$ and $B'=B$. Moreover $V\rho=0$. The $d\times N$ matrix $R=(\rho^{\mathsf T};V)$ therefore satisfies
\begin{equation*}
RR^\dagger=\mathrm{diag}(1,\bar G),\qquad RR^{\mathsf T}=\mathrm{diag}(1,B).
\end{equation*}
With the unitary $Q=(\psi,T)$, define $N$ outcomes $\ket s=Q\,\overline{Re_s}/\|Re_s\|$ with weights $w_s=\|Re_s\|^2$, and add the outcomes $T\ket{f_k}$ with weights $\mu_k$, where $\openone-G=\sum_k\mu_k\ket{f_k}\bra{f_k}$. The first group sums to $Q\,\mathrm{diag}(1,G)\,Q^\dagger$ and the second to $T(\openone-G)T^\dagger$, so the POVM is complete. The first group has real positive overlaps $\braket{s}{\psi}=\rho_s/\|Re_s\|$, and Eq.~\eqref{eq:B:takagi} returns $G_\psi=G$ and $B_\psi=VV^{\mathsf T}=B$. The second group is orthogonal to $\psi$. By Eq.~\eqref{eq:B:F} the POVM attains $F$ with at most $3d-2$ outcomes.

\emph{(c).} On the boundary $G=\openone$. If $B$ is unitary, $F=\frac12(\openone+H_B)$ is a projector of rank $d-1$, so $N=d$ and $RR^\dagger=\openone$. Then $R$ is unitary, all weights equal one, and the outcomes form one orthonormal basis with $|\braket{s}{\psi}|^2=\rho_s^2=1/d$. Any positive unit vector $\rho$ works in the construction, so the overlaps of the basis with $\psi$ can be prescribed at will. Conversely, Proposition~B2 shows that one basis at a regular point gives a projector, so $B$ is unitary.

\emph{(b).} Take a Takagi factorization $B=U\,\mathrm{diag}(\cos\theta_k)\,U^{\mathsf T}$ with $\theta_k\in[0,\pi/2]$, and let $W_\pm=U\,\mathrm{diag}(e^{\pm i\theta_k})\,U^{\mathsf T}$. Both are complex symmetric and unitary, and $B=\frac12(W_++W_-)$. By (c) each $W_\pm$ is the matrix $B_\psi$ of one orthonormal basis unbiased to $\psi$. Mixing the two bases with weights $\frac12$ gives a regular point with $G_\psi=\openone$ and $B_\psi=B$, because $B_\psi$ is linear in the weights.

\emph{(d).} For pure-state models the Holevo bound equals the smallest value of $\mathrm{tr}[W I(M)^{-1}]$ over single-copy measurements~\cite{Matsumoto2002}. Every $F$ in the region~\eqref{eq:B:region} lies below the boundary point $F'=F+\frac12(\openone-\hat G)$, which satisfies $F'\succeq F$ and $F'+\mathcal J^{\mathsf T}F'\mathcal J=\openone$. Since $\mathrm{tr}(WF^{-1})$ decreases with $F$, the minimum is reached on the boundary, and (b) realizes it with two bases.
\end{proof}

Theorem~2(a) is the Fisher-information form of the attainable region, whose covariance form follows from Ref.~\cite{Matsumoto2002}. For $B=0$ the two-basis construction reduces to the coin flip between two bases of Ref.~\cite{Li2016}, which realizes the Fisher-symmetric point.

\section{Mutually unbiased bases and qudit Clifford measurements}
\label{app:mub}

This appendix proves Result~3, its extensions to random Clifford measurements, and a consequence for weak near-optimality.

\begin{theorem}{3 (collective blindness)}
(a)~Let bases $a=0,\dots,K-1$ be pairwise mutually unbiased and mixed with weights $\lambda_a$. For each $a$, the state $\psi=(\ket{b_{a,0}}+\ket{b_{a,1}})/\sqrt2$ and the population change $u=(\ket{b_{a,0}}-\ket{b_{a,1}})/\sqrt2$ give $I(u)/J(u)=\lambda_a$. Hence $c\le\min_a\lambda_a\le1/K$.
(b)~Complete MUBs with $\lambda_a=1/(d+1)$ have $c=1/(d+1)$.
(c)~For the standard complete MUBs in odd prime dimension $d$ and the state $\psi=(\ket0+\ket1)/\sqrt2$ of two computational levels, the conjugate phase change retains $I(iu)/J(u)=d/(d+1)$.
(d)~For the same MUBs, there are states with all $d(d+1)$ outcome probabilities positive and a direction with
\begin{equation}
\frac{I(u)}{J(u)}\le\frac{1}{d+1}\Bigl[1+\frac{4}{d^2(1-d^{-6})}\Bigr].
\label{eq:C:fullsupport}
\end{equation}
(e)~For odd prime $d$, the random Clifford measurement of one qudit has the outcome statistics of complete MUBs, so $c=1/(d+1)$.
(f)~For $n$ qudits of prime dimension $p$, the random Clifford measurement retains $I(u)/J(u)=1/(p+1)$ at $\ket0^{\otimes(n-1)}\otimes(\ket0+\ket1)/\sqrt2$ along $\ket0^{\otimes(n-1)}\otimes(\ket0-\ket1)/\sqrt2$, for every $n$.
\end{theorem}

\begin{proof}
\emph{(a).} Use the path $\psi_\theta=\cos(\pi/4-\theta)\ket{b_{a,0}}+\sin(\pi/4-\theta)\ket{b_{a,1}}$, for which $\partial_\theta\psi|_{\theta=0}=u$ and $J=4$. In basis $a$ the two outcomes $\ket{b_{a,0}},\ket{b_{a,1}}$ have probabilities $\lambda_a\cos^2(\pi/4-\theta)$ and $\lambda_a\sin^2(\pi/4-\theta)$, so at $\theta=0$ they contribute $2\times\lambda_a^2/(\lambda_a/2)=4\lambda_a$. The other outcomes of basis $a$ have probability zero along the whole path. An outcome $x$ of another basis $a'$ has $\braket{b_{a',x}}{b_{a,0}}=e^{i\alpha}/\sqrt d$ and $\braket{b_{a',x}}{b_{a,1}}=e^{i\beta}/\sqrt d$, so with $\chi_{a',x}=\beta-\alpha$,
\begin{equation*}
\begin{aligned}
p_{a',x}(\theta)&=\frac{\lambda_{a'}}{d}\bigl|\cos(\tfrac\pi4-\theta)+e^{i\chi_{a',x}}\sin(\tfrac\pi4-\theta)\bigr|^2\\
&=\frac{\lambda_{a'}}{d}\bigl[1+\cos(2\theta)\cos\chi_{a',x}\bigr],
\end{aligned}
\end{equation*}
which is Eq.~\eqref{eq:fringe} and is stationary at $\theta=0$. Hence $I(u)=4\lambda_a$ and $I/J=\lambda_a$. Since $\sum_a\lambda_a=1$, $\min_a\lambda_a\le1/K$.

\emph{(b).} The upper bound is (a) with $K=d+1$, and the lower bound is Theorem~4(a).

\emph{(c).} For prime $d$, the standard complete set consists of the computational basis and the bases $\ket{b_{a,x}}=d^{-1/2}\sum_j\omega^{aj^2+xj}\ket j$, $a,x\in\mathbb Z_d$, $\omega=e^{2\pi i/d}$~\cite{Wootters1989,Durt2010}. For $\psi=(\ket0+\ket1)/\sqrt2$ one finds $\braket{b_{a,x}}{\psi}=(1+\omega^{-(a+x)})/\sqrt{2d}\ne0$, because $-1$ is not a power of $\omega$ for odd $d$. The only dark outcomes are $\ket j$ with $j\ge2$, and $\braket{j}{u}=0$. Hence $\ell(u)=0$, and Theorem~1 gives $I(iu)=J(u)-I(u)=J(u)\,d/(d+1)$. The same argument applies to every relative phase $e^{i\varphi}\ne-\omega^m$ between the two levels.

\emph{(d).} Let $\ket\chi=(d-2)^{-1/2}\sum_{j\ge2}\ket j$, $\varepsilon=d^{-3}$ and
\begin{equation*}
\psi_{\theta}=\sqrt{1-\varepsilon^2}\bigl[\cos(\tfrac\pi4-\theta)\ket0+\sin(\tfrac\pi4-\theta)\ket1\bigr]+\varepsilon\ket\chi .
\end{equation*}
All computational-basis probabilities are positive, and $J=4(1-\varepsilon^2)$. Because the populations of $\ket j$, $j\ge2$, do not depend on $\theta$, the computational basis contributes $I=4\lambda(1-\varepsilon^2)$ with $\lambda=1/(d+1)$. For an outcome $\ket b$ of another basis, the unperturbed amplitude $a_0=(1+\omega^r)/\sqrt{2d}$ obeys $|a_0|=\sqrt{2/d}\,|\cos(\pi r/d)|\ge\sqrt{2/d}\,\sin(\pi/2d)\ge\sqrt2\,d^{-3/2}$, so the perturbed amplitude satisfies $|a|\ge\sqrt{1-\varepsilon^2}\,|a_0|-\varepsilon\ge|a_0|/2$. The unperturbed derivative of the probability vanishes by (a), so only the perturbation contributes, and $|\partial_\theta p_b|\le2\lambda\varepsilon\sqrt{2/d}$. Each such outcome therefore adds at most $(\partial_\theta p_b)^2/p_b\le16\lambda\varepsilon^2d^2$. Summing over the $d^2$ outcomes and dividing by $J$,
\begin{equation*}
\frac{I}{J}\le\lambda+\frac{16\lambda\varepsilon^2d^4}{4(1-\varepsilon^2)}=\frac1{d+1}\Bigl[1+\frac{4}{d^2(1-d^{-6})}\Bigr].
\end{equation*}

\emph{(e).} For odd prime $d$ the $d(d+1)$ single-qudit stabilizer states are the vectors of the standard complete MUBs~\cite{Durt2010}, and the Clifford group acts transitively on them. A uniformly random Clifford $U$ followed by a computational-basis measurement produces the outcome $U^\dagger\ket x$. By transitivity every stabilizer state appears with the same frequency, so the random Clifford measurement is the POVM $\{\frac1{d+1}\ket s\bra s\}$ of complete MUBs. Merging outcomes that are proportional to the same projector does not change Fisher information, so (b) applies.

\emph{(f).} Let $A$ be the first $n-1$ qudits and $B$ the last one. For each $n$-qudit stabilizer state $\ket s$, the conditional vector $(\bra{0\cdots0}_A\otimes\openone_B)\ket s$ is either zero or $c_s\ket{t_s}$ with $\ket{t_s}$ a single-qudit stabilizer state~\cite{Gottesman1998,Hostens2005}. On the model $\ket{0\cdots0}_A\otimes\phi_\theta$ with $\phi_\theta=\cos(\pi/4-\theta)\ket0+\sin(\pi/4-\theta)\ket1$, each outcome has probability $w|c_s|^2|\braket{t_s}{\phi_\theta}|^2$. Let $\mu_t=\sum_{s:\,t_s=t}w|c_s|^2$. Sandwiching the completeness relation between $\bra{0\cdots0}_A$ and $\ket{0\cdots0}_A$ gives $\sum_t\mu_t\ket t\bra t=\openone_B$. Operators $\openone_A\otimes C$ with $C$ a single-qudit Clifford unitary permute the $n$-qudit stabilizer states and act transitively on the $\ket t$, so $\mu_t$ is constant. Since $\sum_t\ket t\bra t=(p+1)\openone$, $\mu_t=1/(p+1)$. Within each group of outcomes with the same $\ket{t_s}$, the probabilities keep $\theta$-independent proportions, so merging each group leaves the Fisher information unchanged. The measurement thus acts on qudit $B$ as complete MUBs with weights $1/(p+1)$, and (a) gives $I/J=1/(p+1)$.
\end{proof}

For $p=2$, part~(f) gives $1/3$ for every number of qubits, the value that Theorem~8 shows to be exact. For odd $p$, Corollary~G1 combines (f) with a lower bound from shadow estimation.

Theorem~3 also answers, in the negative, the 2-design question that Ref.~\cite{Zhou2026} posed for states of any rank with every parameter unknown. For mixed states with every parameter unknown, single-copy measurements cannot in general retain a constant fraction of the QFI~\cite{Gill2000,Zhou2026}. Reference~\cite{Zhou2026} therefore calls a family of measurements weakly near-optimal on a family of states if at each of these states
\begin{equation}
\begin{aligned}
\mathrm{tr}(JI^{-1})&\le c_1\inf_{\tilde M}\mathrm{tr}(JI_{\tilde M}^{-1}),\\
I&\succeq c_2\,\frac{\mathrm{tr}(J^{-1}I)}{m}\,J,
\end{aligned}
\label{eq:C:weak}
\end{equation}
with constants $c_1,c_2$ independent of the state and the dimension, where $m$ is the number of parameters and the infimum runs over all single-copy measurements $\tilde M$. The same reference asks whether 2-designs are weakly near-optimal for rank-$r$ states with all $r(2d-r)-1$ parameters unknown, a family whose rank-one members are the pure states.

\begin{corollary}{C1 (full-parameter pure states)}
Complete MUBs and SIC measurements are not weakly near-optimal for pure states with all $2(d-1)$ parameters unknown.
\end{corollary}

\begin{proof}
Suppose a measurement satisfies Eq.~\eqref{eq:C:weak} at a pure state. Its Fisher information matrix $I$ is then invertible, since otherwise the first condition fails. A Haar-random basis gives $I_{\mathrm{Haar}}=J/2$ at every pure state~\cite{Hayashi1998,Li2016}, so the first condition gives $\mathrm{tr}(JI^{-1})\le2c_1m$. The Cauchy--Schwarz inequality $\mathrm{tr}(J^{-1}I)\,\mathrm{tr}(JI^{-1})\ge m^2$ and the second condition then give $I\succeq[c_2/(2c_1)]J$, as in Ref.~\cite{Zhou2026}. For complete MUBs this contradicts $I(u)/J(u)=1/(d+1)$ at the encoded qubit of Theorem~3(a) once $d+1>2c_1/c_2$. For SICs it contradicts the value $(1-1/\sqrt{d+1})/d$ of Theorem~4(c) in large dimensions.
\end{proof}

\section{Bounds for 2-designs, SICs and the qubit}
\label{app:extremal}

This appendix proves Result~4. Three elementary facts drive the proofs, namely an $\ell^1$ bound on Fisher information, the zero-sum rule within one basis and the second-moment identity of 2-designs. The second-moment identity also bounds the number of bases from below (Lemma~D4). Throughout, $A=\ket u\bra\psi+\ket\psi\bra u$ with $u\perp\psi$, so that $\mathrm{tr}A=0$, $\mathrm{tr}A^2=2\|u\|^2$ and $\bra sA\ket s=\partial_uq_s$.

\begin{lemma}{D1 ($\ell^1$ bound)}
For one orthonormal basis, $I_B(u)\ge\bigl(\sum_x|\partial_uq_x|\bigr)^2$.
\end{lemma}

\begin{proof}
Dark outcomes have $\partial_uq_x=0$, so the sum runs over outcomes with $q_x>0$. By Cauchy--Schwarz, $\bigl(\sum_x|\partial_uq_x|\bigr)^2=\bigl(\sum_x\frac{|\partial_uq_x|}{\sqrt{q_x}}\sqrt{q_x}\bigr)^2\le\sum_x\frac{(\partial_uq_x)^2}{q_x}\sum_xq_x\le I_B(u)$.
\end{proof}

\begin{lemma}{D2 (zero-sum rule)}
If $v\in\mathbb R^d$ and $\sum_xv_x=0$, then $\|v\|_1^2\ge2\|v\|_2^2$, with equality if and only if $v$ has exactly two nonzero entries.
\end{lemma}

\begin{proof}
Let $P=\sum_{v_x>0}v_x=\sum_{v_x<0}|v_x|$. Then $\|v\|_1=2P$, while $\sum_{v_x>0}v_x^2\le P^2$ and $\sum_{v_x<0}v_x^2\le P^2$, with equality only when each sign class has a single nonzero entry.
\end{proof}

\begin{lemma}{D3 (second moments)}
Every 2-design POVM satisfies $\sum_sw_s\bra sA\ket s^2=2\|u\|^2/(d+1)$.
\end{lemma}

\begin{proof}
By Eq.~\eqref{eq:A:design}, $\sum_sw_s\bra sA\ket s^2=\mathrm{tr}\bigl[\tfrac{2}{d+1}P_{\mathrm{sym}}A\otimes A\bigr]=\frac{(\mathrm{tr}A)^2+\mathrm{tr}A^2}{d+1}=\frac{2\|u\|^2}{d+1}$.
\end{proof}

\begin{lemma}{D4 (number of bases)}
A 2-design built from $K$ orthonormal bases has $K\ge d+1$, with equality only for $d+1$ mutually unbiased bases with equal weights~\cite{Roy2007}.
\end{lemma}

\begin{proof}
For a 2-design, $\sum_{s,t}w_sw_t|\braket{s}{t}|^4=\mathrm{tr}\bigl[(2P_{\mathrm{sym}}/(d+1))^2\bigr]=2d/(d+1)$. For a union of bases with weights $\lambda_a$, pairs within one basis contribute $d\sum_a\lambda_a^2$. For two different bases, Cauchy--Schwarz gives $\sum_{x,y}|\braket{b_{a,x}}{b_{b,y}}|^4\ge\bigl(\sum_{x,y}|\braket{b_{a,x}}{b_{b,y}}|^2\bigr)^2/d^2=1$. Hence $2d/(d+1)\ge d\sum_a\lambda_a^2+\sum_{a\ne b}\lambda_a\lambda_b=(d-1)\sum_a\lambda_a^2+1$, that is, $\sum_a\lambda_a^2\le1/(d+1)$. Since $\sum_a\lambda_a^2\ge1/K$, $K\ge d+1$. Equality forces equal weights and $|\braket{b_{a,x}}{b_{b,y}}|^2=1/d$ for all cross pairs.
\end{proof}

\begin{theorem}{4 (extremality and universality)}
(a)~Every 2-design built from orthonormal bases satisfies $I\succeq J/(d+1)$, and complete MUBs attain the bound.
(b)~Every 2-design satisfies $I\succeq J/[2(d+1)]$.
(c)~Every SIC measurement has $c_{\mathrm{SIC}}=(1-1/\sqrt{d+1})/d$, attained by the in-phase superposition of two SIC vectors along its population change.
(d)~For a qubit, the measurement of the three Pauli bases satisfies $1/3\le I(u)/J(u)\le2/3$ for all states and directions.
\end{theorem}

\begin{proof}
\emph{(a).} The Fisher information of the randomized measurement is $I=\sum_a\lambda_aI_a$, where $I_a$ is the Fisher information of basis $a$ alone. In each basis the responses $\partial_uq_{a,x}=\bra{b_{a,x}}A\ket{b_{a,x}}$ sum to $\mathrm{tr}A=0$. Lemmas~D1 and~D2 give $I_a\ge2\sum_x\bra{b_{a,x}}A\ket{b_{a,x}}^2$, and Lemma~D3 gives
\begin{equation*}
I(u)\ge2\sum_{a,x}\lambda_a\bra{b_{a,x}}A\ket{b_{a,x}}^2=\frac{4\|u\|^2}{d+1}=\frac{J(u)}{d+1}.
\end{equation*}
Equality requires each basis to move at most two outcomes (Lemma~D2) with $|\partial_uq_x|/q_x$ constant on the possible outcomes (Lemma~D1). The encoded qubit of Theorem~3 meets both conditions, because the which-level basis has $\partial_uq=(1,-1,0,\dots)$ at $q=(\frac12,\frac12,0,\dots)$ and every other basis has $\partial_uq=0$.

\emph{(b).} For a 2-design POVM, $(\partial_up_s)^2/p_s=w_s\bra sA\ket s^2/q_s\ge w_s\bra sA\ket s^2$ because $q_s\le1$. Dark outcomes contribute zero on both sides. Lemma~D3 gives $I(u)\ge2\|u\|^2/(d+1)=J(u)/[2(d+1)]$.

\emph{(c).} Let $\{\ket{s_j}\}_{j=1}^{d^2}$ be a SIC, $w=1/d$, $r=1/\sqrt{d+1}$, $a_j=\bra{s_j}A\ket{s_j}$, $q_0=(1+r)/2$ and $m=2(1-r^2)=2d/(d+1)$. Lemma~D3 gives $\sum_ja_j^2=m\|u\|^2$. Cauchy--Schwarz over the possible outcomes gives
\begin{equation}
I(u)=\frac1d\sum_j\frac{a_j^2}{q_j}\ge\frac{\bigl(\frac1d\sum_ja_j^2\bigr)^2}{\frac1d\sum_jq_ja_j^2},
\label{eq:D:cs}
\end{equation}
so it suffices to show
\begin{equation}
\sum_jq_ja_j^2\le q_0\,m\,\|u\|^2 .
\label{eq:D:claim}
\end{equation}
Indeed, Eqs.~\eqref{eq:D:cs} and~\eqref{eq:D:claim} give $I(u)\ge m\|u\|^2/(dq_0)=4(1-r)\|u\|^2/d$, that is, $I/J\ge(1-r)/d$.

If all $q_j\le q_0$, Eq.~\eqref{eq:D:claim} follows from $\sum_ja_j^2=m\|u\|^2$. Otherwise let $Q=q_1>q_0$ be the largest overlap and write $r=\cos\vartheta$ and $Q=\cos^2x$. Since $q_0=\cos^2(\vartheta/2)$, $0\le x<\vartheta/2$. For $j\ne1$ the Fubini--Study angle between $s_j$ and $\psi$ is at least $\vartheta-x$, so $q_j\le h:=\cos^2(\vartheta-x)\le q_0$. Since $u\perp\psi$, $|\braket{s_1}{u}|^2\le(1-Q)\|u\|^2$ and $a_1^2\le4Q(1-Q)\|u\|^2$. Therefore
\begin{equation*}
\begin{aligned}
\sum_jq_ja_j^2&\le Qa_1^2+h\sum_{j\ne1}a_j^2\\
&\le\bigl[hm+4Q(1-Q)(Q-h)\bigr]\|u\|^2 .
\end{aligned}
\end{equation*}
With $y=\vartheta/2-x>0$ one has $q_0-h=\sin(\vartheta+y)\sin y$, $Q-h=2\sin\vartheta\sin y\cos y$, $4Q(1-Q)=\sin^2(2x)$ and $m=2\sin^2\vartheta$, so
\begin{equation*}
\begin{aligned}
&q_0m-hm-4Q(1-Q)(Q-h)\\
&\quad=2\sin\vartheta\sin y\bigl[\sin\vartheta\sin(\vartheta+y)-\sin^2(2x)\cos y\bigr]\ge0,
\end{aligned}
\end{equation*}
because $\sin(\vartheta+y)\ge\sin\vartheta\cos y$ and $\sin(2x)\le\sin\vartheta$ for $0\le2x\le\vartheta\le\pi/2$. This proves Eq.~\eqref{eq:D:claim}.

For the equality, rephase two SIC vectors so that $\braket{s_1}{s_2}=r>0$ and take $\psi=(s_1+s_2)/\sqrt{2(1+r)}$ and $u=(-s_1+s_2)/\sqrt{2(1+r)}$. Every other SIC vector overlaps $s_1$ and $s_2$ with the same modulus $r$, so, as in Theorem~3(a), its probability is stationary and $a_j=0$. The two remaining outcomes have $q_1=q_2=q_0$ and $a_1^2=a_2^2=(1-r)^2$, while $J=4(1-r)/(1+r)$. Hence $I/J=(1-r)/d$.

\emph{(d).} Write the qubit state through its Bloch vector $\boldsymbol r$ with $|\boldsymbol r|=1$, so that $J=|\dot{\boldsymbol r}|^2$. The Pauli basis along axis $k$ gives $I_k=\dot r_k^2/(1-r_k^2)$, with $I_k=0$ when $|r_k|=1$. Since $1-r_k^2\le1$, $I=\frac13\sum_kI_k\ge\frac13|\dot{\boldsymbol r}|^2=J/3$. At regular points Theorem~1 then gives $I(u)=J(u)-I(iu)\le2J/3$. At an eigenstate of a Pauli operator, say $\boldsymbol r=(0,0,1)$, one finds $I_z=0$ and $I_x+I_y=J$, so $I/J=1/3$ in every direction.
\end{proof}

The factor of two between parts~(a) and~(b) is the zero-sum rule of Lemma~D2, by which a basis that responds to a change moves at least two of its outcomes. SICs carry no basis structure, and part~(c) shows that their exact value lies strictly between the two bounds. Part~(d) shows that the bound $(d+2)/[4(d+1)]$ of Ref.~\cite{Zhou2026} is tight at $d=2$, where it equals $1/3$.

\section{Averages and typical states}
\label{app:typical}

This appendix proves the cancellation identity~\eqref{eq:denominator} and Result~5. The proof of Result~5 makes the harmonic picture of Sec.~\ref{sec:averages} precise.

\subsection{The cancellation identity}

\begin{lemma}{E1 (cancellation identity)}
For every 2-design POVM, every $\psi$ and all $x,y\perp\psi$,
\begin{equation}
\sum_sw_s\,\braket{\psi}{s}^2\,\braket{s}{x}\braket{s}{y}=0 .
\label{eq:E:cancel}
\end{equation}
In particular $\sum_sw_sz_s^2=0$ for every direction $u$, while Theorem~1 gives $\eta(u)\|u\|^2=\mathrm{Re}\sum_{s:\,q_s>0}w_sz_s^2/q_s$.
\end{lemma}

\begin{proof}
By Eq.~\eqref{eq:A:design}, the left side equals $\bra{\psi\psi}\tfrac{2}{d+1}P_{\mathrm{sym}}\ket{xy}=\tfrac{2}{d+1}\braket{\psi}{x}\braket{\psi}{y}=0$.
\end{proof}

In the matrix language of Eq.~\eqref{eq:B:takagi}, Lemma~E1 states that $\sum_sw_s\braket{\psi}{s}^2\,\bar v_sv_s^\dagger=0$, whereas $B_\psi$ is the same sum with every term divided by $q_s$. Second moments control the first sum, while Fisher information uses the second.

\subsection{Typical states}

For complete MUBs and a pure state $\psi$, let $c(\psi)=\min_uI(u)/J(u)$ be the worst-direction fraction. A Haar-random state is regular for any finite POVM with probability one, so Proposition~B2 gives $c(\psi)=\frac12(1-\|B_\psi\|)$ almost surely, where $\|\cdot\|$ is the operator norm.

\begin{theorem}{5 (typical states)}
Let $M$ be a complete set of MUBs in dimension $d$ and $\psi$ Haar random. Then
(a)~$\mathbb E\,c(\psi)\ge\frac12(1-\sqrt{\mu_d})>\frac12(1-1/\sqrt2)$ for every $d$, where
\begin{equation}
\mu_d=\frac{d^2h_d-1}{d+1},\qquad h_d=1+(d-1)\ln\Bigl(1-\frac1d\Bigr);
\label{eq:E:mu}
\end{equation}
(b)~$c(\psi)\to\frac12$ in probability as $d\to\infty$ through prime powers.
\end{theorem}

The proofs use Gaussian vectors. Let $g\in\mathbb C^d$ have independent standard complex Gaussian entries, $\mathbb E|g_j|^2=1$, so that $\psi=g/\|g\|$ is Haar random. For a measurement vector $s$, the amplitude $\zeta_s=\braket{s}{g}$ is a standard complex Gaussian, and $\mathbb E\,\zeta_s\bar\zeta_t=\braket{s}{t}$. Normalization does not change phases, so $e^{-2i\arg\braket{s}{\psi}}=e^{-2i\arg\zeta_s}$.

\begin{lemma}{E2 (phase harmonics)}
For a standard complex Gaussian $\zeta$ and $n\ge1$ let
\begin{equation}
\Phi_n(\zeta)=\frac{\bar\zeta^{\,2}L^{(2)}_{n-1}(|\zeta|^2)}{\sqrt{n(n+1)}},
\label{eq:E:harmonics}
\end{equation}
with $L^{(2)}_{n-1}$ a generalized Laguerre polynomial. Then
\begin{equation}
\begin{aligned}
e^{-2i\arg\zeta}&=\sum_{n\ge1}\frac{\Phi_n(\zeta)}{\sqrt{n(n+1)}}\quad\text{in }L^2,\\
\mathbb E\,\Phi_n(\zeta_1)\overline{\Phi_m(\zeta_2)}&=\delta_{nm}\,\bar\rho^{\,n+1}\rho^{\,n-1}
\end{aligned}
\label{eq:E:expansion}
\end{equation}
for jointly Gaussian standard $\zeta_1,\zeta_2$ with $\mathbb E\,\zeta_1\bar\zeta_2=\rho$. In particular
\begin{equation}
\mathbb E\,e^{-2i\arg\zeta_1}e^{2i\arg\zeta_2}=\frac{\bar\rho^{\,2}}{|\rho|^2}\,H(|\rho|),
\label{eq:E:twopoint}
\end{equation}
with
\begin{equation*}
H(r)=\sum_{n\ge1}\frac{r^{2n}}{n(n+1)}=1+\frac{1-r^2}{r^2}\ln(1-r^2).
\end{equation*}
\end{lemma}

\begin{proof}
Up to normalization and sign, $\Phi_n$ is the complex Hermite polynomial $H_{n-1,n+1}(\zeta,\bar\zeta)$. These polynomials are orthogonal, and their correlations for jointly Gaussian arguments follow from the complex Mehler formula~\cite{Janson1997}, which gives the second relation in Eq.~\eqref{eq:E:expansion}. The function $e^{-2i\arg\zeta}=\bar\zeta^2/|\zeta|^2$ has angular momentum $-2$, so it lies in the closed span of the $\Phi_n$. Its coefficients are $\mathbb E\,\overline{\Phi_n(\zeta)}\,\bar\zeta^2/|\zeta|^2=\int_0^\infty e^{-t}\,t\,L^{(2)}_{n-1}(t)\,dt/\sqrt{n(n+1)}=1/\sqrt{n(n+1)}$, because $L^{(2)}_{n-1}=\sum_{j<n}L^{(1)}_j$ and the $L^{(1)}_j$ are orthogonal for the weight $te^{-t}$. Equation~\eqref{eq:E:twopoint} follows by summing the series.
\end{proof}

\begin{proof}[\textit{Proof of Theorem~5(a).}]
Lift $B_\psi$ to the full space as $C_\psi=\sum_sw\,e^{-2i\arg\braket{s}{\psi}}\ket{\bar s}\bra s$ with $w=1/(d+1)$. Completeness gives $C_\psi\ket\psi=\ket{\bar\psi}$, and in the frame $(\psi,T)$ one finds $(\psi,T)^{\mathsf T}C_\psi(\psi,T)=\mathrm{diag}(1,B_\psi)$, so $\|C_\psi\|_F^2=1+\|B_\psi\|_F^2$. Expanding,
\begin{equation*}
\|C_\psi\|_F^2=\sum_{s,t}w^2\,e^{-2i(\arg\zeta_s-\arg\zeta_t)}\braket{s}{t}^2 .
\end{equation*}
The $d(d+1)$ diagonal terms give $d/(d+1)$. Distinct vectors of one basis are orthogonal. For vectors of different bases $|\braket{s}{t}|^2=1/d$, and Eq.~\eqref{eq:E:twopoint} turns each of the $d^3(d+1)$ ordered pairs into $w^2h_d/d$, with $h_d=H(d^{-1/2})$. Hence $\mathbb E\|C_\psi\|_F^2=[d+d^2h_d]/(d+1)$ and $\mathbb E\|B_\psi\|_F^2=\mu_d$. Since $1/[n(n+1)]\le\frac12$, $h_d\le\sum_n d^{-n}/2=1/[2(d-1)]$, which gives $\mu_d<\frac12$. Jensen's inequality and $\|B_\psi\|\le\|B_\psi\|_F$ then give $\mathbb E\,c(\psi)=\frac12(1-\mathbb E\|B_\psi\|)\ge\frac12(1-\sqrt{\mu_d})$.
\end{proof}

For part~(b) split $C_\psi$ into harmonics, $C_\psi=\sum_{n\ge1}C_n$ with
\begin{equation}
C_n=\frac{w}{\sqrt{n(n+1)}}\sum_s\Phi_n(\zeta_s)\ket{\bar s}\bra s,\qquad B_n=T^{\mathsf T}C_nT .
\label{eq:E:layers}
\end{equation}

\begin{lemma}{E3 (the lowest harmonic cancels)}
$B_1=0$.
\end{lemma}

\begin{proof}
$\Phi_1(\zeta)=\bar\zeta^2/\sqrt2$, so $C_1=\frac w2\sum_s\braket{g}{s}^2\ket{\bar s}\bra s$, and $B_1$ has entries $\frac w2\sum_s\braket{g}{s}^2\braket{s}{x}\braket{s}{y}$ with $x,y\in\psi^\perp$. These vanish by Lemma~E1, because $g\propto\psi$.
\end{proof}

\begin{lemma}{E4 (the higher harmonics average out)}
(i)~$\mathbb E\|\sum_{n>m}B_n\|_F^2\le1/(m+1)$ for every $m\ge1$.
(ii)~For every fixed $n\ge2$ there is $\gamma_n<\infty$, independent of $d$, with $\mathbb E\,\mathrm{tr}(B_nB_n^\dagger)^2\le\gamma_n/d$.
\end{lemma}

\begin{proof}
(i)~Different harmonics are uncorrelated by Eq.~\eqref{eq:E:expansion}. For one harmonic, the same computation as in part~(a) gives
\begin{equation*}
\begin{aligned}
\mathbb E\|C_n\|_F^2&=\frac{w^2}{n(n+1)}\sum_{s,t}|\braket{s}{t}|^{2n+2}\\
&=\frac{1}{n(n+1)}\,\frac{d+d^{2-n}}{d+1}\le\frac{1}{n(n+1)}\qquad(n\ge2).
\end{aligned}
\end{equation*}
Compression by $T$ does not increase the Frobenius norm, and $\sum_{n>m}1/[n(n+1)]=1/(m+1)$.

(ii)~Write $A_s=\ket{\bar s}\bra s$ and expand $\mathrm{tr}(C_nC_n^\dagger C_nC_n^\dagger)$ as a sum over quadruples of measurement vectors, weighted by $w^4$, by the product of four harmonics, and by a trace $\mathrm{tr}(A_iA_j^\dagger A_kA_l^\dagger)$. The trace is bounded by the product of the moduli of the overlaps around the cycle $i\to j\to k\to l\to i$, and every step between distinct vectors costs at most $d^{-1/2}$. Steps between distinct vectors of one basis give zero. The expectation of the four harmonics is a sum over Wick pairings with no pairing inside one harmonic~\cite{Janson1997}. Every pairing between distinct vectors again costs at most $d^{-1/2}$, and each $\Phi_n$ carries $2n$ Gaussian factors. Sort the quadruples by the number $\nu$ of distinct vectors. For $\nu=1$ there are $O(d^2)$ quadruples with bounded terms. For $\nu=2$ there are $O(d^4)$ quadruples, and the cycle has at least two steps between distinct vectors, so the trace is $O(d^{-1})$. For $\nu=3$ there are $O(d^6)$ quadruples, the trace is $O(d^{-3/2})$, and a vector that occurs once forces at least four pairings between distinct vectors, so the expectation of the four harmonics is $O(d^{-2})$. For $\nu=4$ there are $O(d^8)$ quadruples, the trace is $O(d^{-2})$, and all $4n$ pairings join distinct vectors, giving $O(d^{-2n})$. With $w^4\le d^{-4}$ the four classes contribute $O(d^{-2})$, $O(d^{-1})$, $O(d^{-3/2})$ and $O(d^{2-2n})$. Compression by $T$ does not increase Schatten norms.
\end{proof}

\begin{proof}[\textit{Proof of Theorem~5(b).}]
Fix $\varepsilon,\delta>0$ and choose $m$ with $4/[(m+1)\varepsilon^2]<\delta/2$. By Lemma~E4(i) and Markov's inequality, $\Pr(\|\sum_{n>m}B_n\|>\varepsilon/2)<\delta/2$. By Lemma~E4(ii), $\mathbb E\|\sum_{n=2}^mB_n\|_{S_4}^4\le m^3\sum_{n=2}^m\gamma_n/d$, so $\Pr(\|\sum_{n=2}^mB_n\|>\varepsilon/2)<\delta/2$ for all large $d$. By Lemma~E3, $B_\psi=\sum_{n\ge2}B_n$, hence $\Pr(\|B_\psi\|>\varepsilon)<\delta$ for all large $d$, and $c(\psi)=\frac12(1-\|B_\psi\|)\to\frac12$ in probability.
\end{proof}

The proof does not track the rate. Numerically the mean of $c(\psi)$ behaves as $\frac12-0.65/\sqrt d$ (Fig.~\ref{fig:typical}), and $\mathbb E\,\mathrm{tr}(B_\psi B_\psi^\dagger)^2=1/(2d)+O(d^{-2})$, consistent with $\|B_\psi\|=\Theta(d^{-1/2})$. The same proof applies to SICs, whose $d^2$ vectors overlap pairwise with modulus $(d+1)^{-1/2}$.

\section{Random bases}
\label{app:random}

This appendix proves Result~6. The bases $B_1,\dots,B_K$ are independent and Haar random, $B_a=\{U_a\ket{e_x}\}_x$, and the measurement mixes them with weights $1/K$. We work on
\begin{equation}
\begin{aligned}
\Omega&=\{(\psi,u):\|\psi\|=\|u\|=1,\ \braket{\psi}{u}=0\},\\
\varrho&=\|\psi-\psi'\|+\|u-u'\|,
\end{aligned}
\label{eq:F:omega}
\end{equation}
where $\varrho$ is the distance between $(\psi,u)$ and $(\psi',u')$, and $J=4$. For a basis $B$ define the $\ell^1$ response
\begin{equation}
Y_B(\psi,u)=\sum_x|\partial_uq_x|=2\sum_x\bigl|\mathrm{Re}\,\braket{x}{u}\braket{\psi}{x}\bigr|\in[0,2].
\label{eq:F:Y}
\end{equation}
Lemma~D1 gives $I_B(u)\ge Y_B^2$, and convexity gives
\begin{equation}
I(u)=\frac1K\sum_aI_{B_a}(u)\ge\Bigl(\frac1K\sum_aY_{B_a}(\psi,u)\Bigr)^2 .
\label{eq:F:jensen}
\end{equation}
The same $\ell^1$ quantity and its Haar mean (Lemma~F2) appear in Ref.~\cite{Voroninski2013}, where a constant number of Haar bases is shown to determine pure states stably. A small-ball estimate (Lemma~F3) on a net (Lemmas~F1 and F4) makes the number of bases explicit.

\begin{lemma}{F1 (Lipschitz)}
$|Y_B(\psi,u)-Y_B(\psi',u')|\le2\varrho\bigl((\psi,u),(\psi',u')\bigr)$ for every basis $B$.
\end{lemma}

\begin{proof}
$\mathrm{Re}\braket{x}{u}\braket{\psi}{x}-\mathrm{Re}\braket{x}{u'}\braket{\psi'}{x}=\mathrm{Re}\braket{x}{u}\braket{\psi-\psi'}{x}+\mathrm{Re}\braket{x}{u-u'}\braket{\psi'}{x}$. Summing the moduli over an orthonormal basis and applying Cauchy--Schwarz gives at most $\|u\|\|\psi-\psi'\|+\|u-u'\|\|\psi'\|$.
\end{proof}

\begin{lemma}{F2 (Haar mean)}
$\mathbb E\,Y_B(\psi,u)=1$ for every $(\psi,u)\in\Omega$ and every $d\ge2$.
\end{lemma}

\begin{proof}
For each $x$, the pair $(\braket{\psi}{x},\overline{\braket{x}{u}})$ has the law of two coordinates $(v_1,v_2)$ of a uniform unit vector of $\mathbb C^d$. Write $v=g/\|g\|$ with $g$ standard complex Gaussian, $\mathbb E|g_j|^2=1$. The norm $\|g\|$ is independent of $v$, so $\mathbb E|\mathrm{Re}\,v_1\bar v_2|=\mathbb E|\mathrm{Re}\,g_1\bar g_2|/\mathbb E\|g\|^2$. With $g_j=a_j+ib_j$ and $a_j,b_j\sim\mathcal N(0,\frac12)$, $\mathrm{Re}\,g_1\bar g_2=a_1a_2+b_1b_2$ has characteristic function $(1+t^2/4)^{-1}$, a Laplace law with mean absolute value $\frac12$. Hence $\mathbb E\,Y_B=2d\cdot\frac{1}{2d}=1$.
\end{proof}

\begin{lemma}{F3 (small ball)}
For $d\ge3$ and $t\in(0,1]$,
\begin{equation}
\Pr\bigl[Y_B(\psi,u)\le t\bigr]\le3.2\,d^{3/2}\ln(e/t)\,(C_0t)^{d-1},
\label{eq:F:smallball}
\end{equation}
where $C_0=4\sqrt e$.
\end{lemma}

\begin{proof}
The pair $(a,b)=(U^\dagger\psi,U^\dagger u)$ is a uniformly random orthonormal pair in $\mathbb C^d$, and $Y=2\sum_x|\mathrm{Re}\,\bar a_xb_x|$.

\emph{Step 1 (remove phases).} Write $a_x=r_xe^{i\theta_x}$ and $D=\mathrm{diag}(e^{-i\theta_x})$. Then $Da=r$ is real and nonnegative, $Db$ is uniform on the unit sphere of $r^\perp$, and $\mathrm{Re}\,\bar a_xb_x=r_x\mathrm{Re}(Db)_x$. We may thus take $a=r$ and $b=\xi+i\zeta$ uniform on the unit sphere of $V\oplus iV$, where $V=\{v\in\mathbb R^d:\ r\cdot v=0\}$ has dimension $d-1$. Then $Y=2\sum_xr_x|\xi_x|$, and $r_x>0$ almost surely.

\emph{Step 2 (separate the length).} Let $\varrho_\xi=\|\xi\|$ and $\hat\xi=\xi/\varrho_\xi$. Then $\varrho_\xi^2\sim\mathrm{Beta}(\alpha,\alpha)$ with $\alpha=(d-1)/2$, and $\hat\xi$ is uniform on the unit sphere of $V$ and independent of $\varrho_\xi$. We have $Y=2\varrho_\xi S(\hat\xi)$ with the norm $S(v)=\sum_xr_x|v_x|$ on $V$.

\emph{Step 3 (sphere to Gaussian).} Let $G$ be a standard Gaussian vector on $V$, so that $G=\|G\|\hat\xi'$ with $\hat\xi'$ uniform and independent of $\|G\|$. Since $\Pr[\|G\|\le2\sqrt{d-1}]\ge\frac34$, we get $\Pr[S(\hat\xi)\le s]\le\frac43\Pr[S(G)\le2\sqrt{d-1}\,s]$.

\emph{Step 4 (density times volume).} The density of $G$ on $V$ is at most $(2\pi)^{-(d-1)/2}$, so $\Pr[S(G)\le\sigma]\le(2\pi)^{-(d-1)/2}\sigma^{d-1}\mathrm{vol}_{d-1}\{v\in V:S(v)\le1\}$. The map $\Lambda=\mathrm{diag}(1/r_x)$ sends the section $B_1^d\cap\mathbb 1^\perp$ of the cross-polytope onto the unit ball $\{S\le1\}$ of $S$, and comparing volumes gives
\begin{equation*}
\mathrm{vol}_{d-1}\{S\le1\}=\frac{\mathrm{vol}_{d-1}(B_1^d\cap\mathbb 1^\perp)}{\sqrt d\,X}\le\frac{2^{d-1}}{(d-1)!\,X},
\end{equation*}
with $X=\prod_xr_x$,
because the orthogonal projection of the section onto a coordinate hyperplane lies in $B_1^{d-1}$ and scales volumes by $1/\sqrt d$. With $(d-1)!\ge((d-1)/e)^{d-1}$, Steps~3 and~4 give
\begin{equation*}
\Pr[S(\hat\xi)\le s\mid r]\le\frac{2\,(c_2s)^{d-1}}{(d-1)^{(d-1)/2}X},\qquad c_2=2e\sqrt{2/\pi}.
\end{equation*}

\emph{Step 5 (average over the amplitudes).} The probabilities $r_x^2$ are uniform on the simplex, and the Dirichlet integral gives $M_d=\mathbb E[1/X]=(d-1)!\,\pi^{d/2}/\Gamma(d/2)$. Stirling-type bounds give $M_d(d-1)^{-(d-1)/2}\le2.1\sqrt d\,(2\pi/e)^{d/2}$, and the AM--GM inequality gives $M_d\ge d^{d/2}$.

\emph{Step 6 (average over the length).} Since $Y\le t$ means $S(\hat\xi)\le t/(2\varrho_\xi)$, Steps~4 and~5 give $\Pr[Y\le t]\le\mathbb E\min(1,\kappa_t\varrho_\xi^{-(d-1)})$ with $\kappa_t=2(c_2t/2)^{d-1}M_d(d-1)^{-(d-1)/2}$. The density of $\varrho_\xi$ is at most $2\varrho^{d-2}/B(\alpha,\alpha)$, and $1/B(\alpha,\alpha)\le\alpha2^{2\alpha-1}$. Splitting the integral at $\varrho_0=\kappa_t^{1/(d-1)}$ gives
\begin{equation*}
\begin{aligned}
\Pr[Y\le t]&\le\kappa_t2^{d-2}\Bigl[1+\ln_+\frac1{\kappa_t}\Bigr]\\
&\le2.1\sqrt{2\pi/e}\,\sqrt d\,(C_0t)^{d-1}\,d\ln\frac et,
\end{aligned}
\end{equation*}
where the last step uses Step~5, $C_0=c_2\sqrt{2\pi/e}=4\sqrt e$, and $1+\ln_+(1/\kappa_t)\le d\ln(e/t)$. Since $2.1\sqrt{2\pi/e}<3.2$, Eq.~\eqref{eq:F:smallball} follows.
\end{proof}

\begin{lemma}{F4 (net)}
For every $\delta\in(0,1]$, $\Omega$ contains a $2\delta$-net for $\varrho$ with at most $(1+4/\delta)^{4d}$ points.
\end{lemma}

\begin{proof}
By volume comparison, the unit sphere of $\mathbb R^{2d}$ has an $\varepsilon$-net with at most $(1+2/\varepsilon)^{2d}$ points. With $\varepsilon=\delta/2$ the product of two such nets is a $\delta$-net of the product of spheres. Replacing each net point whose $\delta$-ball meets $\Omega$ by a point of $\Omega$ in that ball gives a $2\delta$-net of $\Omega$.
\end{proof}

\begin{theorem}{6 (random bases)}
For each $K\ge5$ let $\delta_K=[4\cdot5^4(5C_0K)^K]^{-1/(K-4)}$ and $c_K=\delta_K^2/4$. There is $d_0(K)$ such that for all $d\ge d_0(K)$, with probability at least $1-2^{-d}$ over the bases, $I\succeq c_KJ$ for every pure-state model. If $K(d-1)>4d$, then $c(M)>0$ almost surely. At least three bases are necessary.
\end{theorem}

\begin{proof}
Let $\delta=\delta_K$ and let $\mathcal N$ be the net of Lemma~F4, with $|\mathcal N|\le(5/\delta)^{4d}$. Fix a net point and write $Y_a=Y_{B_a}$. If $\frac1K\sum_aY_a\le5\delta$, then every $Y_a\le5K\delta$. By independence and Lemma~F3, the event that every $Y_a\le5K\delta$ has probability at most $[3.2\,d^{3/2}L_K(5C_0K\delta)^{d-1}]^K$ with $L_K=\ln[e/(5K\delta_K)]$. A union bound over the net gives
\begin{equation*}
\begin{aligned}
&\Pr\Bigl[\exists\,\text{net point with }\tfrac1K\textstyle\sum_aY_a\le5\delta\Bigr]\\
&\quad\le\bigl[5^4(5C_0K)^K\delta^{K-4}\bigr]^d\,\frac{(3.2\,d^{3/2}L_K)^K}{(5C_0K\delta)^K}\\
&\quad=4^{-d}\,\frac{(3.2\,d^{3/2}L_K)^K}{(5C_0K\delta_K)^K},
\end{aligned}
\end{equation*}
by the choice of $\delta_K$. The right side is at most $2^{-d}$ once $2^d\ge[3.2\,d^{3/2}L_K/(5C_0K\delta_K)]^K$, which defines $d_0(K)$. On the event that every net point has $\frac1K\sum_aY_a>5\delta$, every $(\psi,u)\in\Omega$ lies within $2\delta$ of a net point, so Lemma~F1 gives $\frac1K\sum_aY_{B_a}(\psi,u)>5\delta-4\delta=\delta$, and Eq.~\eqref{eq:F:jensen} gives $I(u)>\delta^2=c_KJ$. For general $\delta$ the same union bound is at most $C(K,d)\,\delta^{K(d-1)-4d}$, which tends to zero as $\delta\to0$ when $K(d-1)>4d$, so $c(M)>0$ almost surely.

For the last statement, let $\psi$ be a vector of $B_1$. All other outcomes of $B_1$ are dark and the outcome $\psi$ does not respond, so $B_1$ contributes nothing. A second basis contributes a Fisher matrix of rank at most $d-1$, because its $d$ probabilities sum to one. With $K=2$ some direction $w$ therefore has $I(w)=0$, and Lemma~B3 turns this into a vanishing ratio at nearby points with $\ell=0$.
\end{proof}

\begin{proposition}{F5 (order-$d$ bases)}
If $K\ge570\,d$, then with probability at least $1-e^{-d}$ the measurement satisfies $I\succeq J/16$ for every pure-state model.
\end{proposition}

\begin{proof}
Take the $\frac18$-net of Lemma~F4 with $\delta=\frac1{16}$, which has at most $65^{4d}$ points. At a fixed net point the $Y_a$ are independent, lie in $[0,2]$ and have mean $1$ (Lemma~F2), so Hoeffding's inequality~\cite{Hoeffding1963} gives $\Pr[\frac1K\sum_aY_a\le\frac34]\le e^{-K/32}$. The union bound $65^{4d}e^{-K/32}\le e^{-d}$ holds for $K\ge32(1+4\ln65)\,d\approx566.4\,d$. On the event that every net point has $\frac1K\sum_aY_a>\frac34$, Lemma~F1 extends $\frac1K\sum_aY_a\ge\frac12$ to all of $\Omega$, and Eq.~\eqref{eq:F:jensen} gives $I\ge\frac14=J/16$.
\end{proof}

The constants $c_K$ of Theorem~6 are small, for example $c_5\approx3\times10^{-30}$, and they decrease with $K$, because $\Pr[\frac1K\sum_aY_a\le t]\le\Pr[Y\le Kt]^K$ discards most of the averaging. The actual constants are much larger and grow with $K$ (Fig.~\ref{fig:settings} and Appendix~\ref{app:numerics}). Monte Carlo sampling of $2\times10^7$ orthonormal pairs per dimension gives $\Pr[Y\le t]\approx(1.67\,t)^{d-1}$ up to a prefactor, which confirms the exponent $d-1$ of Lemma~F3. The count $K(d-1)>4d$ expresses the same dimension counting that governs the injectivity of generic bases on pure states~\cite{Mondragon2013,Kech2017}.

\section{Relation to shadow estimation}
\label{app:learning}

This appendix proves Result~7 and the consequences used in Sec.~\ref{sec:learning}. Consider a randomized measurement $\{w_s\ket s\bra s\}$ that forms a 2-design. Its measurement channel has the inverse $\mathcal M^{-1}(O)=(d+1)O-\mathrm{tr}(O)\openone$, and an outcome $s$ yields the classical-shadow estimate $\hat o_s=\bra s\mathcal M^{-1}(O)\ket s$ of an observable $O$~\cite{Huang2020shadows}. The squared shadow norm is the largest second moment of this estimate over states,
\begin{equation}
\|O\|_{\mathrm{sh}}^2=\max_\sigma\sum_sw_s\bra s\sigma\ket s\,\hat o_s^{\,2},
\quad
\kappa=\sup_{O\ne0,\ \mathrm{tr}O=0}\frac{\|O\|_{\mathrm{sh}}^2}{\mathrm{tr}\,O^2}.
\label{eq:G:shadow}
\end{equation}
This is the shadow norm of Refs.~\cite{Huang2020shadows,Mao2025} written for the outcome vectors of the measurement. For 3-designs $\kappa\le3$, and for 2-designs $\kappa\le d+1$~\cite{Huang2020shadows,Yang2026}.

\begin{theorem}{7 (correspondence between sensing and learning)}
Every randomized measurement that forms a 2-design satisfies $I(u)\ge J(u)/(2\kappa)$ for all pure states and all directions, that is, $I\succeq J/(2\kappa)$.
\end{theorem}

\begin{proof}
Let $A=\ket u\bra\psi+\ket\psi\bra u$, so that $\partial_up_s=w_s\bra sA\ket s$, $p_s=w_s\bra s\rho\ket s$ with $\rho=\ket\psi\bra\psi$, and dark outcomes have $\bra sA\ket s=0$. Cauchy--Schwarz with the test function $c_s=\bra sA\ket s$ gives
\begin{equation*}
I(u)\ge\frac{\bigl(\sum_sw_s\bra sA\ket s^2\bigr)^2}{\sum_sw_s\bra s\rho\ket s\bra sA\ket s^2}.
\end{equation*}
By Lemma~D3 the numerator equals $4\|u\|^4/(d+1)^2$. Since $A$ is traceless, $\bra sA\ket s=\hat o_s/(d+1)$ for the shadow estimate of $A$, so the denominator is the second moment of that estimate in the state $\rho$, divided by $(d+1)^2$. It is therefore at most $\|A\|_{\mathrm{sh}}^2/(d+1)^2\le\kappa\,\mathrm{tr}A^2/(d+1)^2=2\kappa\|u\|^2/(d+1)^2$. Hence $I(u)\ge2\|u\|^2/\kappa=J(u)/(2\kappa)$.
\end{proof}

The denominator in this proof is the variance of the shadow estimate of the operator $\partial\rho=A$ that carries the signal. Sensing a parameter with a randomized measurement thus costs at most what learning $\partial\rho$ with the same measurement costs. This Cauchy--Schwarz argument also underlies Refs.~\cite{Du2026,Cho2026} and the local shadow estimator of Ref.~\cite{Zhou2026}. For 2-designs, $\kappa\le d+1$ reproduces Theorem~4(b).

\begin{remark}
\textit{(Third moments.)} For a 3-design the denominator can be computed exactly. Equation~\eqref{eq:A:design} with $t=3$ gives $\sum_sw_s\bra s\rho\ket s\bra sA\ket s^2=\frac{6}{(d+1)(d+2)}\mathrm{tr}[P^{(3)}_{\mathrm{sym}}A\otimes A\otimes\rho]=\frac{4\|u\|^2}{(d+1)(d+2)}$, because $\mathrm{tr}[P^{(3)}_{\mathrm{sym}}A\otimes A\otimes\rho]=\frac16[\mathrm{tr}A^2+2\,\mathrm{tr}(A^2\rho)]=\frac23\|u\|^2$. The Cauchy--Schwarz bound in the proof of Theorem~7 then becomes $I\succeq\frac{d+2}{4(d+1)}J$, the constant of Ref.~\cite{Zhou2026}.
\end{remark}

\begin{corollary}{G1 (Clifford measurements of qudits)}
For $n$ qudits of odd prime dimension $p$, random Clifford measurements satisfy $1/(4p-2)\le c\le1/(p+1)$ for every $n$.
\end{corollary}

\begin{proof}
For traceless $O$, Ref.~\cite{Mao2025} gives $\|O\|_{\mathrm{sh}}^2\le(2p-3)\,\mathrm{tr}O^2+2\|O\|_{\mathrm{op}}^2\le(2p-1)\,\mathrm{tr}O^2$, so $\kappa\le2p-1$ and Theorem~7 gives the lower bound. The upper bound is Theorem~3(f).
\end{proof}

For one qudit the correspondence is tight on the encoded qubit of Theorem~3. There $A=\ket0\bra0-\ket1\bra1$ is diagonal in a stabilizer basis, and Ref.~\cite{Mao2025} gives $\|A\|_{\mathrm{sh}}^2=(d+1)\|A\|_{\mathrm{op}}^2=d+1$. The proof of Theorem~7, with $\|A\|_{\mathrm{sh}}^2$ kept in place of $\kappa\,\mathrm{tr}A^2$, then gives $I\ge J/(d+1)$, which is the exact value.

\begin{corollary}{G2 (one magic gate)}
Measure in the Clifford orbit of a magic state, a measurement that a random Clifford circuit followed by a single $T$ gate and a Fourier gate realizes~\cite{Mao2025}. This measurement satisfies $I\succeq J/15$ in every dimension.
\end{corollary}

\begin{proof}
The Clifford group is a unitary 2-design, so every Clifford orbit is a 2-design and Theorem~7 applies. The shadow norm of every magic orbit obeys $\kappa\le15/2$ in the normalization of Eq.~\eqref{eq:G:shadow}~\cite{Zhu2024magic}.
\end{proof}

\begin{corollary}{G3 (number of bases)}
Five random bases give near-optimal sensing of all pure states in every large dimension (Theorem~6). Worst-case-optimal shadow estimation requires of order $d^2$ bases~\cite{Yang2026}.
\end{corollary}

\begin{proof}
A measurement with $N$ outcome vectors has a worst-case shadow norm of at least $(d^2-1)^2/(Nd)$~\cite{Yang2026}. A bounded $\kappa$ therefore needs $N=\Omega(d^3)$ outcome vectors, that is, $\Omega(d^2)$ bases, and order $d^2$ bases suffice~\cite{Yang2026}.
\end{proof}

Sensing can be near-optimal even where learning is costly, since five random bases have $\kappa$ of order $d^2$ and still retain a dimension-independent fraction.

\section{Random Clifford measurements of qubits}
\label{app:clifford}

This appendix proves Result~8. Let $D=2^n$, let $\mathcal P_n$ be the $D^2$ Hermitian Pauli operators without phases, and let the measurement be the uniform mixture of all stabilizer bases, that is, the POVM of all $N=D\prod_{j=1}^n(2^j+1)$ stabilizer states with weights $w=D/N$. For a pure state write $r_P=\bra\psi P\ket\psi$. Pauli completeness gives $\sum_{P\in\mathcal P_n}r_P^2=D$.

\begin{theorem}{8 (random Clifford measurements of qubits)}
For every $n$, $I\succeq J/3$ for all pure-state models, and $c=1/3$. At every stabilizer state the normalized Fisher matrix equals $\openone/3$. Along suitable paths of regular states that approach a stabilizer state, the worst direction retains $I/J\to1/3$ and its conjugate $I/J\to2/3$.
\end{theorem}

The proof has three parts, namely the value at stabilizer states, a recursion over Pauli observables and an inequality that closes the recursion.

\subsection{Stabilizer states}

\begin{proposition}{H1 (isotropy at stabilizer states)}
At every stabilizer state, $J^{-1/2}IJ^{-1/2}=\openone/3$.
\end{proposition}

\begin{proof}
By Clifford covariance we may take $\psi=\ket{0^n}$. The tangent space splits into the planes $P_x=\mathrm{span}_{\mathbb R}\{\ket x,i\ket x\}$, $x\ne0$. The phase gate on qubit $j$ fixes $\psi$, rotates the planes with $x_j=1$ by $90^\circ$ and fixes the others. Invariance of the Fisher form under this gate removes the cross terms between planes that differ in bit $j$, and makes the form on each plane commute with a $90^\circ$ rotation, hence scalar. The CNOT gates fix $\psi$ and act on the labels $x\ne0$ as $\mathrm{GL}(n,2)$, which is transitive, so all planes carry the same scalar $c_0$. The trace then fixes $c_0$. Summing Eq.~\eqref{eq:B:budget} over a real orthonormal basis of the tangent space gives $\mathrm{tr}(J^{-1}I)=(D-1)-wN_\perp$, where $N_\perp$ counts stabilizer states orthogonal to $\psi$. A stabilizer state is supported on an affine subspace with amplitudes of equal modulus~\cite{Dehaene2003}, so it overlaps $\ket{0^n}$ exactly when its support is a linear subspace. There are $\binom nk_2$ linear subspaces of dimension $k$ and $2^{k(k+3)/2}$ stabilizer states on each, so the Cauchy binomial theorem gives
\begin{equation*}
\begin{aligned}
N-N_\perp&=\sum_{k=0}^n\binom nk_2\,2^{k(k-1)/2}\,4^k\\
&=\prod_{j=2}^{n+1}(1+2^j)=N\,\frac{2D+1}{3D}.
\end{aligned}
\end{equation*}
Hence $wN_\perp=(D-1)/3$, $\mathrm{tr}(J^{-1}I)=\frac23(D-1)$, and $c_0=\frac13$.
\end{proof}

\subsection{Recursion over Pauli observables}

The stabilizer bases correspond to maximal isotropic subspaces $L$ of $\mathbb F_2^{2n}$, and each $L$ contains $D-1$ nonidentity Pauli operators. Drawing $L$ uniformly is therefore the same as drawing a nonidentity Pauli $P$ uniformly and then drawing $L\ni P$ uniformly. The second step draws a uniform maximal isotropic subspace of $P^\perp/\langle P\rangle\cong\mathbb F_2^{2(n-1)}$. The measurement can thus be performed sequentially, by measuring $P$, with eigenprojectors $\Pi_\pm$, and then measuring each nonzero branch in a uniformly random stabilizer basis of the $(n-1)$-qubit code space $\Pi_\pm$.

Fix a unit direction $u\perp\psi$, so that $J=4$. For the two-outcome measurement of $P$ let $p_\pm=\bra\psi\Pi_\pm\ket\psi$ and $c_\pm=\bra\psi\Pi_\pm\ket u$. Its Fisher information along $u$ and $iu$, and the QFI lost in dark branches, are
\begin{equation}
\begin{aligned}
K_P(u)&=4\sum_{\pm:\,p_\pm>0}\frac{(\mathrm{Re}\,c_\pm)^2}{p_\pm},\\
K_P(iu)&=4\sum_{\pm:\,p_\pm>0}\frac{(\mathrm{Im}\,c_\pm)^2}{p_\pm},\\
Z_P(u)&=4\sum_{\pm:\,p_\pm=0}\|\Pi_\pm u\|^2 .
\end{aligned}
\label{eq:H:coarse}
\end{equation}
We write $K_n$ and $Z_n$ for their averages over the $D^2-1$ nonidentity Pauli operators.

\begin{lemma}{H2 (chain rule)}
For a branch with $p_\pm>0$, the conditional state $\psi_\pm=\Pi_\pm\psi/\sqrt{p_\pm}$ moves in the direction $w_\pm=\Pi_\pm u/\sqrt{p_\pm}-(c_\pm/p_\pm)\psi_\pm$, with QFI $J_\pm=4\|w_\pm\|^2$, and
\begin{equation}
\sum_{\pm:\,p_\pm>0}p_\pm J_\pm=4-Z_P(u)-K_P(u)-K_P(iu).
\label{eq:H:chain}
\end{equation}
The Fisher information of the sequential measurement is $K_P(u)+\sum_\pm p_\pm I_\pm$, where $I_\pm$ is the Fisher information of the branch measurement on $(\psi_\pm,w_\pm)$.
\end{lemma}

\begin{proof}
$p_\pm J_\pm=4(\|\Pi_\pm u\|^2-|c_\pm|^2/p_\pm)$. Summing over the nonzero branches and using $\|\Pi_+u\|^2+\|\Pi_-u\|^2=1$ gives Eq.~\eqref{eq:H:chain}. The second statement is the chain rule of classical Fisher information for $p(\pm,y)=p_\pm\,p(y\mid\pm)$.
\end{proof}

\begin{proposition}{H3 (closing condition)}
If $K_n(iu)+Z_n(u)\le2K_n(u)$ for all $n$, $\psi$ and $u$, then $I\ge J/3$ for all $n$.
\end{proposition}

\begin{proof}
Induction on $n$. For $n=1$ the stabilizer bases are the three Pauli bases, and Theorem~4(d) gives $I\ge J/3$. For $n\ge2$, fix the first Pauli $P$. Each branch measurement is a uniform stabilizer measurement of $n-1$ qubits, so the induction hypothesis gives $I_\pm\ge J_\pm/3$, and Lemma~H2 gives $I\ge K_P(u)+\frac13[4-Z_P(u)-K_P(u)-K_P(iu)]$. Averaging over $P$,
\begin{equation*}
I\ge\frac{4-Z_n(u)+2K_n(u)-K_n(iu)}{3}\ge\frac43=\frac J3 .
\end{equation*}
Only the marginal law of the stabilizer basis drawn in each branch enters, so correlations between the bases of the two branches do not matter.
\end{proof}

\subsection{Closing the recursion}

It remains to verify the closing condition of Proposition~H3. For a nonidentity $P$ write $x_P=\partial_ur_P=2\,\mathrm{Re}\bra\psi P\ket u$, $z_P=r_P^2$ and $w_P=r_Px_P=\frac12\partial_uz_P$. At points with $|r_P|<1$ the measurement of $P$ has Fisher information $F_P(u)=x_P^2/(1-r_P^2)$.

\begin{lemma}{H4 (anticommutation budget)}
For every nonidentity $P$, with $N(P)$ the Pauli operators that anticommute with $P$,
\begin{equation}
\sum_{Q\in N(P)}r_Q^2=\frac D2\bigl(1-r_P^2\bigr),\qquad \sum_{Q\in N(P)}w_Q=-\frac D2\,w_P .
\label{eq:H:budget}
\end{equation}
\end{lemma}

\begin{proof}
Since $\psi$ is pure, $r_P^2=\mathrm{tr}(\rho P\rho P)$. Expanding $\rho=D^{-1}\sum_Qr_QQ$ and using $PQP=\pm Q$ gives $Dr_P^2=\sum_{Q:[P,Q]=0}r_Q^2-\sum_{Q\in N(P)}r_Q^2$. Together with $\sum_Qr_Q^2=D$ this is the first identity, and differentiating along $u$ gives the second.
\end{proof}

\begin{lemma}{H5 (nearly definite observables commute)}
Let $\mathcal H=\{P\ne\openone:z_P>\frac12\}$. Any two elements of $\mathcal H$ commute, and every Pauli operator anticommutes with at most $D/2$ elements of $\mathcal H$. If $P\in\mathcal H$ anticommutes with $Q$, then $z_P\le1-z_Q$.
\end{lemma}

\begin{proof}
Two anticommuting Hermitian involutions act as two orthogonal Bloch axes of a qubit, so $r_P^2+r_Q^2\le1$, which gives the first and last statements. The labels of $\mathcal H$ in $\mathbb F_2^{2n}$ span an isotropic subspace $S$ with $|S|\le D$. For any $Q$, the symplectic form with $Q$ is a linear functional on $S$, which vanishes either on all of $S$ or on exactly half of it.
\end{proof}

\begin{lemma}{H6 (information exclusion)}
At points with $0<z_P<1$ for all nonidentity $P$,
\begin{equation}
\sum_{P\ne\openone}\frac{w_P^2}{1-z_P}\le\sum_{P\ne\openone}\frac{w_P^2}{z_P}.
\label{eq:H:exclusion}
\end{equation}
\end{lemma}

\begin{proof}
With $f(z)=(2z-1)/z$, the claim is $\sum_P\frac{f(z_P)}{1-z_P}w_P^2\le0$. Split the sum into $\mathcal H$ and $\mathcal L=\{P:z_P<\frac12\}$. Terms with $z_P=\frac12$ vanish. For $P\in\mathcal H$, Lemma~H4 and Cauchy--Schwarz give
\begin{equation*}
\begin{aligned}
w_P^2&=\frac4{D^2}\Bigl(\sum_{Q\in N(P)}w_Q\Bigr)^2\le\frac4{D^2}\sum_{Q\in N(P)}z_Q\sum_{Q\in N(P)}\frac{w_Q^2}{z_Q}\\
&=\frac{2(1-z_P)}{D}\sum_{Q\in N(P)}\frac{w_Q^2}{z_Q}.
\end{aligned}
\end{equation*}
By Lemma~H5 every $Q\in N(P)$ lies in $\mathcal L$. Multiplying by $f(z_P)/(1-z_P)>0$ and summing over $\mathcal H$,
\begin{equation*}
\begin{aligned}
\sum_{P\in\mathcal H}\frac{f(z_P)\,w_P^2}{1-z_P}&\le\sum_{Q\in\mathcal L}\frac{w_Q^2}{z_Q}\,\frac2D\sum_{P\in\mathcal H\cap N(Q)}f(z_P)\\
&\le\sum_{Q\in\mathcal L}\frac{w_Q^2}{z_Q}\,\frac{1-2z_Q}{1-z_Q},
\end{aligned}
\end{equation*}
where the last step uses that $f$ increases, $z_P\le1-z_Q$ and $|\mathcal H\cap N(Q)|\le D/2$ (Lemma~H5). The right side equals $-\sum_{Q\in\mathcal L}f(z_Q)w_Q^2/(1-z_Q)$, which is the claim.
\end{proof}

\begin{lemma}{H7 (two-sided window)}
For every $\psi$ and unit $u\perp\psi$,
\begin{equation}
\frac{2D}{D^2-1}\le K_n(u),\qquad K_n(u)+Z_n(u)\le\frac{4D}{D^2-1}.
\label{eq:H:window}
\end{equation}
\end{lemma}

\begin{proof}
Pauli completeness gives $\sum_{P\ne\openone}x_P^2=D\,\mathrm{tr}A^2=2D$ for $A=\ket u\bra\psi+\ket\psi\bra u$. For the lower bound, $F_P(u)\ge x_P^2$, and dark branches occur only when $|r_P|=1$, where $x_P=0$. Averaging gives $K_n(u)\ge2D/(D^2-1)$. For the upper bound at points with $0<z_P<1$, $w_P^2/z_P=x_P^2$ and $w_P^2/(1-z_P)=r_P^2x_P^2/(1-r_P^2)$, so Lemma~H6 gives $\sum_PF_P(u)=\sum_Px_P^2+\sum_Pr_P^2x_P^2/(1-r_P^2)\le2\sum_Px_P^2=4D$. At other points, move along $\psi_\epsilon=\cos\epsilon\,\psi+\sin\epsilon\,u$ with a small generic perturbation that makes all $0<z_P<1$, apply the bound, and let the perturbation and $\epsilon$ tend to zero. Terms with $|r_P|<1$ converge to $F_P(u)$, and a term with $|r_P|=1$ converges to $4\|\Pi_-u\|^2=Z_P(u)$, the QFI of the dark branch. Hence $\sum_P[F_P(u)+Z_P(u)]\le4D$, and averaging gives the upper bound.
\end{proof}

\begin{proof}[\textit{Proof of Theorem~8.}]
Applying the upper bound of Lemma~H7 to $iu$ and using $Z_n(iu)=Z_n(u)$ gives $K_n(iu)+Z_n(u)\le4D/(D^2-1)\le2K_n(u)$. Proposition~H3 then gives $I\succeq J/3$, and Proposition~H1 shows that the bound is attained. For the last statement, let $\psi_0$ be a stabilizer state and $w$ a unit direction with $\braket{s}{w}\ne0$ for every dark outcome $s$, so that the path of Lemma~B3 runs through regular states. By Proposition~H1, $I(w)/J=I(iw)/J=\frac13$ at $\psi_0$, so Eq.~\eqref{eq:B:budget} gives $\eta_0(w)=0$ and $\ell_0(w)=\frac13$. Along the path of Lemma~B3, the direction $w_\varepsilon$ retains $\frac12[1-\ell_0+\eta_0]\to\frac13$, and the budget gives $\frac23$ to its conjugate.
\end{proof}

\begin{remark}
At a stabilizer state the Fisher information divides each squared response by the overlap $q_s$, and the overlaps of the responding outcomes follow an exact law. For $\psi=\ket{0^n}$ and $u=\ket x$, weighting each outcome by $w\bra sA\ket s^2$ gives $\Pr(q_s=2^{-k})=\binom{n-1}{k-1}_2\,2^{k(k-1)/2}/\prod_{j=1}^{n-1}(1+2^j)$ for $k=1,\dots,n$, and $\mathbb E(1/q_s)=2(D+1)/3$. With $\sum_sw\bra sA\ket s^2=2/(D+1)$, this reproduces $I/J=\frac13$.
\end{remark}

For odd prime $p$ the same recursion, with the $p$ eigenspaces of a Pauli line in place of $\Pi_\pm$, reduces the conjectured value $c=1/(p+1)$ of $n$-qudit Clifford measurements to the single condition $K_n(iu)+Z_n(u)\le pK_n(u)$. Theorem~3(f) shows that this value would be tight.

\section{Numerical methods}
\label{app:numerics}

This appendix describes how the numerical values in Figs.~\ref{fig:price}--\ref{fig:finite} and Table~\ref{tab:A1} were obtained.

\subsection{Fisher ratios and adversarial searches}

Every measurement set is checked for completeness and for the design property it is meant to have, to a precision of $10^{-14}$. For a state $\psi$ and a direction $u$, the ratio $I(u)/J(u)$ follows from Eq.~\eqref{eq:A:fisher}, and the worst-direction fraction at a regular point follows from Proposition~B2 as $\frac12(1-\sigma_{\max})$. Outcomes with $|\braket{s}{\psi}|<10^{-6}$ are treated as dark. The threshold $10^{-6}$ sits well above the precision of the numerically constructed SICs.

Worst cases are estimated by two independent adversaries. Adversary~A minimizes $\frac12(1-\sigma_{\max})$ over $\psi$ with Nelder--Mead from random and structured starting points (vectors of one basis and two-component superpositions within and across bases), followed by a gradient polish. Adversary~B minimizes the directional ratio $I(u)/J(u)$ jointly over $(\psi,u)$ with L-BFGS-B and uses no eigendecomposition. Both return upper estimates of the retained fraction. On complete MUBs, run without the known worst state as a start, both recover $1/(d+1)$ in every dimension tested.

For $K=4,5,6,8,12$ random bases and $d=5,7,11,17,23$, the searches give minima of about $10^{-4}$, $0.004$, $0.01$, $0.04$ and $0.09$, with no trend in $d$, and scans with $K=d+1,2d,3d,4d,6d,8d$ for $d=3$--$17$ fall on the same curve (Fig.~\ref{fig:settings}). Two complete MUB sets related by a Haar-random unitary give $0.22$--$0.27$ for $d=3$--$17$. For Weyl--Heisenberg SICs in $d=3$--$8$, three searches with $40$--$200$ restarts all stop at $(1-1/\sqrt{d+1})/d$. Random Clifford measurements give $0.333333$ for two and three qubits, $0.250000$ for two qutrits and $1/6$ for two ququints, the last two equal to $1/(p+1)$. The Clifford orbit of a magic state gives $0.388$, $0.394$, $0.326$ and $0.441$ for $d=3,5,7,11$. For Haar-random states and complete MUBs, Fig.~\ref{fig:typical} uses 400 samples for $d\le19$ and 200 samples for $d\ge23$.

\subsection{A certified lower bound for twelve random bases}

For $d=3$ and three fixed draws of $K=12$ bases we certify the worst case by interval branch and bound. The state space $\mathbb{CP}^{2}$ is covered by coordinate charts $\tilde\psi=e_j+\sum_{i\ne j}z_ie_i$ with $|z_i|\le1$, and boxes are cubes in the real coordinates of $z$. Over a box with center $\tilde\psi_0$, each amplitude moves by at most $R_s=\sqrt2\,h\sum_{i\ne j}|\braket{s}{e_i}|$ for half-width $h$. If $R_s<|\braket{s}{\tilde\psi_0}|/2$, the phase factor $e^{-2i\arg\braket{s}{\psi}}$ moves by at most $\delta_s=2R_s/|\braket{s}{\tilde\psi_0}|$, and outcome $s$ is called clean. For the remaining outcomes the phase is replaced by its worst value on a grid of 24 or 48 points, with the grid error added to the bound. The direction $u$ is orthogonal to $\psi$ and only approximately orthogonal to the box center. This approximate orthogonality costs a term of order $\varrho=2\sqrt{2(d-1)}\,h/\|\tilde\psi_0\|$. Together the amplitude and phase bounds, the grid error and the orthogonality term give, for every state in the box and every unit $u\perp\psi$,
\begin{equation}
\frac{I(u)}{J(u)}\ge\frac12\bigl(1-\beta_{\mathrm{box}}-\varepsilon_{\mathrm{box}}\bigr)-10^{-9},
\label{eq:I:box}
\end{equation}
where $\beta_{\mathrm{box}}$ bounds the Takagi norm of the clean part plus the phase grid of the remaining outcomes, $\varepsilon_{\mathrm{box}}=\lambda_{\max}\bigl(\sum_s w\delta_s\ket s\bra s\bigr)$ over clean outcomes, and $10^{-9}$ absorbs floating-point error. Boxes whose bound exceeds the value to be certified are discarded, and the others are halved until none remain. The resulting lower bounds are $0.1583$, $0.1019$ and $0.1051$, against search values of $0.17584$, $0.11316$ and $0.11062$, that is, 90\%, 90\% and 95\% of the search values, using $3.4$--$6.7\times10^7$ boxes. A soundness check compares Eq.~\eqref{eq:I:box} with exact values at 300 random points in each of 1869 random boxes. A third of these 1869 boxes sit on zeros of one or two amplitudes, and the smallest margin is $2.9\times10^{-4}$. In this dimension the searches therefore come within 10\% of the true worst case.

\subsection{Finite-sample simulations}

Figure~\ref{fig:finite} uses one-parameter models along geodesics, $\psi_\theta=\cos\theta\,\psi_0+\sin\theta\,v_0$ with $v_0\perp\psi_0$ a unit vector and true value $\theta=0$, so $J=4$ and the quantum Cram\'er--Rao bound is $1/(4N)$. The measurements are complete MUBs for odd prime $d$, twelve Haar-random bases with seed $1000+d$, and the Clifford orbit of $T^\dagger F^\dagger\ket0$ with the canonical $T$ gate. The orbit has 72, 600, 784, 14520, 9464 and 83232 elements for $d=3,5,7,11,13,17$. The second moments of the twelve random bases deviate from those of a 2-design by $0.02$--$0.09$. Each simulated experiment draws $N$ outcomes, including the basis label, and estimates $\theta$ by maximum likelihood. Outcomes with identical probability functions are merged first, and the estimator runs a global search on 360 grid points in $[-\pi/2,\pi/2)$ followed by 12 Newton steps. Each data point averages 4000 experiments, and error bars are standard errors. The typical state is the median, by worst-direction fraction, of 21 Haar-random states with seed $2000+d$, and it moves along its worst direction.

The predicted cost $J/I$ uses the pathwise Fisher information of Appendix~\ref{app:notation}. The pathwise value equals the pointwise value in all cases except the magic orbit at $d=5$, where $\psi_0$ is orthogonal to ten orbit vectors. The ten corresponding outcomes become possible at second order along the path, so the absence of their clicks carries information. The pointwise ratio is $0.489$ and the pathwise ratio $0.523$. The simulation gives $J/I=1.87\pm0.04$, against the pathwise prediction $1.91$. In the encoded-qubit example of Theorem~3 the dark outcomes $\ket j$, $j\ge2$, are orthogonal to the entire path and carry no information under either convention.

At large $d$ the population estimate of the encoded qubit under complete MUBs approaches its asymptotic cost slowly. At $N=10^4$ the relative deficit below $d+1$ is about $0.5\%$, $4.5\%$ and $9\%$ for $d=11,13,17$, each with a standard error of about $2\%$, consistent with roughly $0.2\,d^3/N$. At $N=10^5$ the deficit falls below $1\%$. The cause is a few nearly dark interference outcomes. In odd prime dimension some unbiased outcomes have relative phases within $\pi/d$ of $\pi$, so their probability at equal populations is about $\pi^2/(2d^4)$. A tilt changes the visibility at second order, which changes these small probabilities by a large relative amount, and the estimator gains information of order $\theta^4$ that pulls it toward equal populations. The asymptotic cost $d+1$ is reached once $N\gtrsim d^3$.

The dependence on the operating point in Fig.~\ref{fig:finite}(c) follows from the fringe picture of Sec.~\ref{sec:mub}. At population contrast $t$, each unbiased basis contributes a Fisher fraction that tends to $t/(1+t)$ as $d\to\infty$, so the cost of complete MUBs approaches $(1+t)/t$. The plateau at $d+1$ has a width of order $1/d$.

\bibliography{refs}

\end{document}